\documentclass{article}
\usepackage{ijcai25}

\usepackage{times}
\usepackage{soul}
\usepackage{url}
\usepackage[hidelinks]{hyperref}
\usepackage[utf8]{inputenc}
\usepackage[small]{caption}
\usepackage{graphicx}
\usepackage{amsmath}
\usepackage{amsthm}
\usepackage{booktabs}
\usepackage{algorithm}
\usepackage{algorithmic}
\usepackage[switch]{lineno}
\usepackage[utf8]{inputenc} 
\usepackage[T1]{fontenc}    
\usepackage{url}            
\usepackage{booktabs}       
\usepackage{amsfonts}       
\usepackage{nicefrac}       
\usepackage{microtype}      
\usepackage{xcolor}         
\usepackage{graphicx}
\usepackage{amsmath}
\usepackage{bm}
\usepackage{subfig}
\usepackage{booktabs}
\usepackage{makecell}
\usepackage{multicol}
\usepackage{multirow}
\usepackage{wrapfig}
\usepackage{graphicx} 
\usepackage{subfiles}

\newcommand{\methodname}{scSiameseClu}

\title{scSiameseClu: A Siamese Clustering Framework for Interpreting \\ Single-cell RNA Sequencing Data}

\author{
Ping Xu$^{1,2}$\and
Zhiyuan Ning$^{1,2}$\and
Pengjiang Li$^{1,2}$\and
Wenhao Liu$^3$\and
Pengyang Wang$^{4}$\footnotemark[1]\and \\
Jiaxu Cui$^{5}$\and
Yuanchun Zhou$^{1,2}$\And
Pengfei Wang$^{1,2}$\footnotemark[1]\\
\affiliations
$^1$Computer Network Information Center, Chinese Academy of Sciences, Beijing\\
$^2$University of Chinese Academy of Sciences, Beijing\\
$^3$College of Life Science, Northeast Agricultural University, Harbin\\
$^4$SKL-IOTSC and Department of CIS, University of Macau, Macau \\
$^5$Jilin University, Changchun\\
\emails
\{xuping, ningzhiyuan, pjli\}@cnic.cn, 
b230901013@neau.edu.cn, \\
pywang@um.edu.mo, 
cjx@jlu.edu.cn, 
\{zyc,pfwang\}@cnic.cn
}

\begin{document}

\maketitle
\setcounter{footnote}{0}
\renewcommand{\thefootnote}
{\fnsymbol{footnote}}
\footnotetext[1]{Corresponding authors.}
\footnotetext[2]{Code and datasets are all available at the link: \url{https://github.com/XPgogogo/scSiameseClu}.}

\begin{abstract}
    Single-cell RNA sequencing (scRNA-seq) reveals cell heterogeneity, with cell clustering playing a key role in identifying cell types and marker genes. 
    Recent advances, especially graph neural networks (GNNs)-based methods, have significantly improved clustering performance. 
    However, the analysis of scRNA-seq data remains challenging due to noise, sparsity, and high dimensionality. 
    Compounding these challenges, GNNs often suffer from over-smoothing, limiting their ability to capture complex biological information. 
    In response, we propose \textbf{\methodname}, a novel \textbf{Siamese} \textbf{Clu}stering framework for interpreting \textbf{s}ingle-\textbf{c}ell RNA-seq data, comprising of 3 key steps:
    (1) Dual Augmentation Module, which applies biologically informed perturbations to the gene expression matrix and cell graph relationships to enhance representation robustness; 
    (2) Siamese Fusion Module, which combines cross-correlation refinement and adaptive information fusion to capture complex cellular relationships while mitigating over-smoothing; 
    and (3) Optimal Transport Clustering, which utilizes Sinkhorn distance to efficiently align cluster assignments with predefined proportions while maintaining balance. 
    Comprehensive evaluations on seven real-world datasets demonstrate that~\methodname~outperforms state-of-the-art methods in single-cell clustering, cell type annotation, and cell type classification, providing a powerful tool for scRNA-seq data interpretation.
\end{abstract}

\begin{figure}[!t]
\centering
\subfloat[scNAME]{
\includegraphics[width=0.185\textwidth]{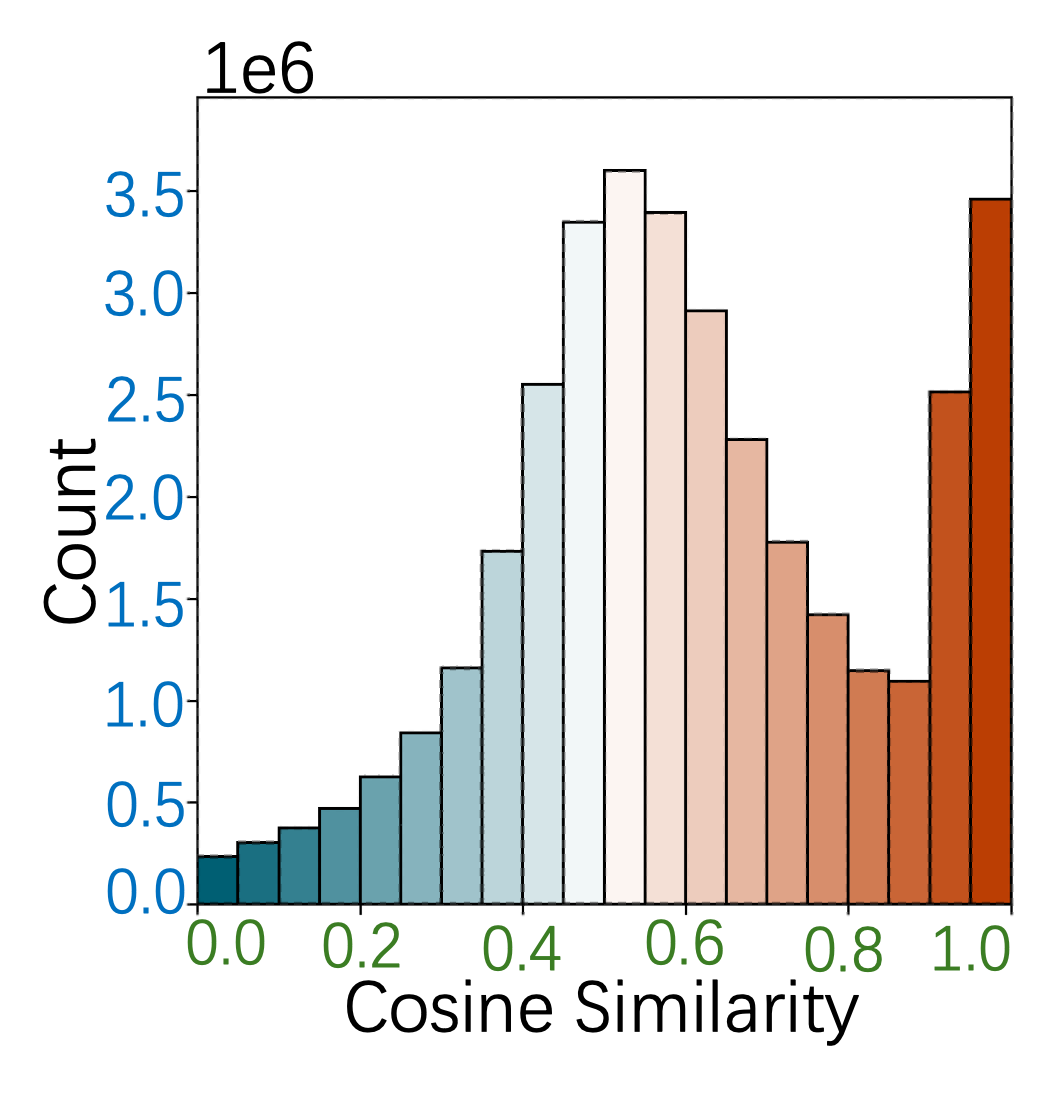}
}
\subfloat[scGNN]{
\includegraphics[width=0.176\textwidth]{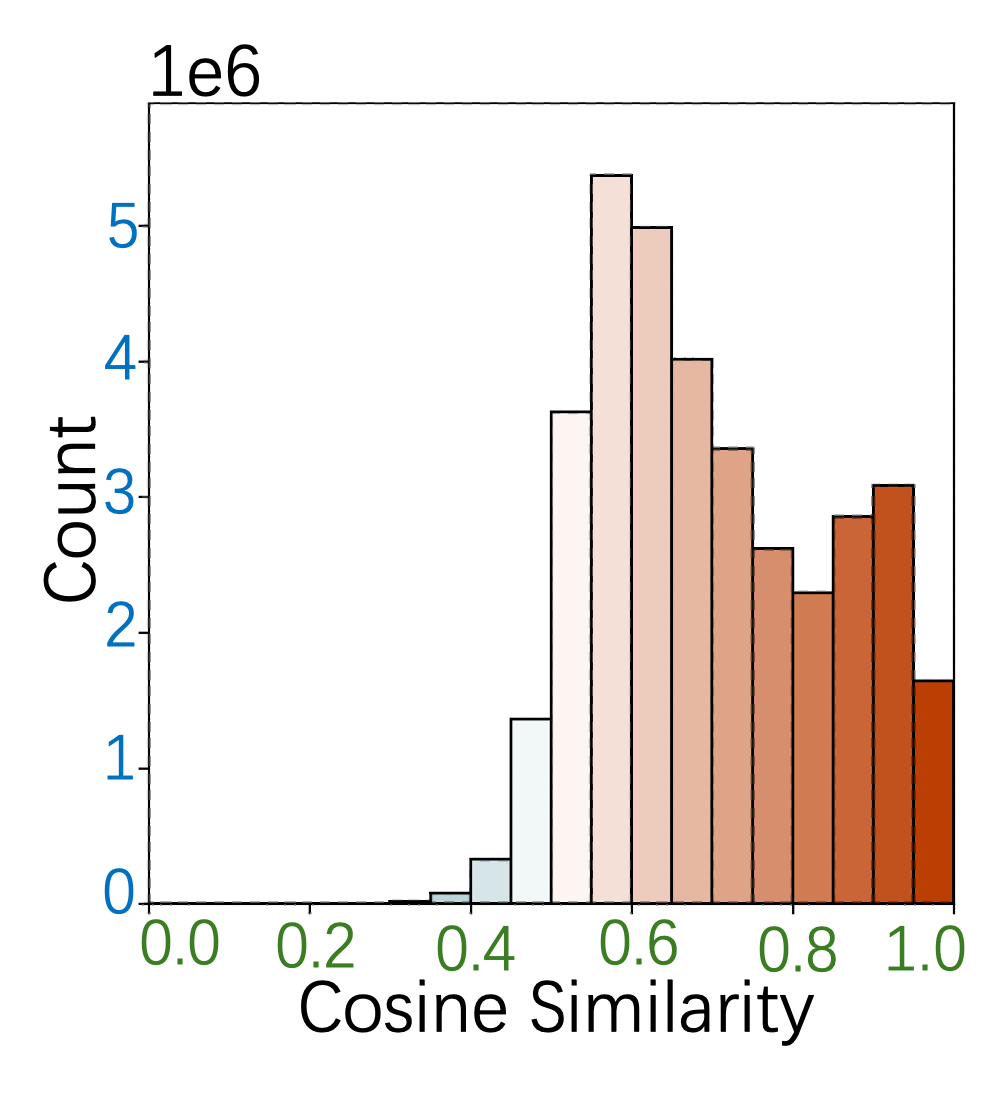}
}
\caption{Similarity distributions of cell embeddings learned by scNAME and scGNN on dataset \emph{Human liver cells}.}
\label{fig:intro_heatmap}
\end{figure}

\section{Introduction}
Single-cell RNA sequencing (scRNA-seq) technology represents a significant advancement in bioinformatics, enabling the capture of comprehensive genetic information from individual cells~\cite{shapiro2013single,wang2025sccompass}. 
Cell clustering, a key step in single-cell RNA sequencing analysis, groups cells by their gene expression patterns to uncover the complex characteristics of distinct cell populations and provide insights into their biological functions and interactions~\cite{kiselev2019challenges,yang2024genecompass}.
Moreover, clustering gene expression patterns further contributes to various downstream tasks, such as marker gene identification and cell type annotation~\cite{xu2025soft}. 
Cluster analysis of scRNA-seq data has been a vibrant research area over the past decade. 

In recent years, advanced computational methods have been increasingly explored to address the challenges of analyzing high-dimensional and sparse scRNA-seq data. 
Classical clustering methods, such as K-means, are straightforward and computationally efficient but struggle to capture the complex, nonlinear relationships in scRNA-seq data. 
Meanwhile, recent research has focused on applying deep learning frameworks for learning representations of scRNA-seq data and classifying samples into distinct clusters. 
A commonly used method is self-supervised learning methods, which uncover effective representations of scRNA-seq data by reconstructing the original input data~\cite{eraslan2019single,lopez2018deep,tian2019clustering}. 
However, these methods primarily focus on extracting features from individual cells, overlooking the complex relationships between cells that are essential for understanding cellular diversity. 
Benefiting from the powerful utilization of graph information, graph neural networks (GNNs) have been applied to analyze scRNA-seq data by modeling cells as nodes and their interactions as edges, effectively capturing both gene expressions and cell graphs to comprehensively represent cellular heterogeneity
~\cite{wang2021scgnn,gan2022deep,zhan2023scmic}. 

Though good performance has been achieved, previous GNN-based works still face limitations when tackling the following challenges: 
\textbf{(1) Deficient exploration on intercellular information:} when applying GNN-based models to single-cell data, the graph construction process often overlooks the sparsity and noise inherent in scRNA-seq data, frequently leading to reduced model robustness. 
Most current methods rely on simple topologies, like cosine similarity, to build cell graphs from gene expression matrices~\cite{gan2022deep,WANG2021165},  but do not handle the sparsity and noise inherent in scRNA-seq data. 
This leads to fragile graph representations, weakening GNNs' ability to capture cellular interactions and limiting clustering performance and insights into cellular dynamics.
\textbf{(2) Insufficient prevention of representation collapse in cell embeddings:} 
GNNs-based models for scRNA-seq data frequently encounter representation collapse, where embeddings of biologically distinct cells become overly similar in latent space. 
This issue arises from the inability of current methods to preserve the diversity of cell representations in sparse and noisy scRNA-seq data. 
As shown in Fig.\ref{fig:intro_heatmap}, both the deep learning-based scNAME~\cite{wan2022scname} and the GNN-based scGNN~\cite{wang2021scgnn} exhibit varying degrees of representation collapse. 
The problem is more pronounced in scGNN, where the cosine similarity between almost all cell embeddings exceeds $0.5$, indicating a severe loss of diversity in cell embeddings. 
Such collapse diminishes the discriminative power of embeddings, blurring cluster boundaries and limiting the ability to distinguish different cell populations, ultimately degrading the effectiveness of GNN models in clustering tasks~\cite{zbontar2021barlow}. 

To tackle the aforementioned challenge, we propose \textbf{~\methodname}, a \textbf{Siamese} \textbf{Clu}stering framework for interpreting \textbf{s}ingle-\textbf{c}ell RNA sequencing data. 
\methodname~is designed to capture and refine complex intercellular information while learning discriminative and robust representations across both gene and cellular features. 
\methodname~leverages three key components: dual augmentation module to enrich data, siamese fusion module to preserve critical information and reduce redundancy, and optimal transport clustering to align cluster distributions. 
It effectively explores intricate information, mitigates representation collapse, and achieves clearer cell population separation, excelling in scRNA-seq clustering and other biological tasks. 

Our framework offers the following contributions:
\begin{itemize}
    \item We present~\methodname, a novel Siamese-based clustering framework tailored for scRNA-seq data that captures intricate information from gene expression and cell graphs to learn discriminative and robust cell embeddings, improving clustering outcomes and downstream tasks.
    \item We introduce key components: (i) dual augmentation with distinct noise on gene expression and cell graphs to mitigate dropout effects and improve robustness; (ii) siamese fusion for cross-correlation and adaptive information fusion to enhance robustness; and (iii) optimal transport clustering to align distributions. 
    \item Experimental results on seven datasets demonstrate that~\methodname~outperforms state-of-the-art (SOTA) methods in clustering and other biological tasks.
\end{itemize}

\section{Related Work}
\noindent\textbf{Deep Clustering for scRNA-seq.} 
In scRNA-seq data analysis, early clustering algorithms like Phenograph~\cite{levine2015data}, MAGIC~\cite{van2018recovering}, and Seurat~\cite{wang2025sccompass} utilize k-nearest neighbor (KNN) graphs to model cell relationships, while SIMLR~\cite{wang2018simlr} and MPSSC~\cite{park2018spectral} employ multiple kernel functions to derive robust similarity measures from different data representations. These methods, however, often struggle with high-dimensional data and complex nonlinear features, making them susceptible to noise. 
Deep learning approaches have since become prominent. DCA~\cite{eraslan2019single} employs a Zero-Inflated Negative Binomial (ZINB) autoencoder to model scRNA-seq data distributions, capturing nonlinear gene dependencies. SCVI~\cite{lopez2018deep} and SCVIS~\cite{ding2018interpretable}, which rely on autoencoders for dimensionality reduction, often face over-regularization due to their Gaussian distribution assumption. Additionally, 
scDeepCluster~\cite{tian2019clustering} combines ZINB-based autoencoders with deep embedding clustering to optimize feature learning and clustering.  
Most approaches ignore inter-gene and inter-cell correlations, focusing solely on individual cell expression profiles.

\noindent\textbf{Graph Clustering for scRNA-seq.} 
Recently, graph-based clustering methods have gained considerable attention for their effectiveness~\cite{tu2021deep,liu2022deep,ning2025deep}. 
Methods like scGAE~\cite{luo2021topology} and graph-sc~\cite{ciortan2022gnn} utilize graph autoencoders to embed scRNA-seq data while preserving topological structures. 
scGAC~\cite{cheng2022scgac} constructs cell graphs and employs a self-optimization approach for simultaneous representation learning and clustering. 
Notably, scGNN~\cite{wang2021scgnn} uses GNNs and multi-modal autoencoders to aggregate cell-cell relationships and model gene expression patterns, while scDSC~\cite{gan2022deep} combines a ZINB model-based autoencoder with GNN modules using a mutually supervised strategy. 
These methods often struggle with representation collapse, severely limiting their ability to accurately model cell relationships.

\begin{figure*}[thbp]
    \centering
    \includegraphics[width=0.9\linewidth]{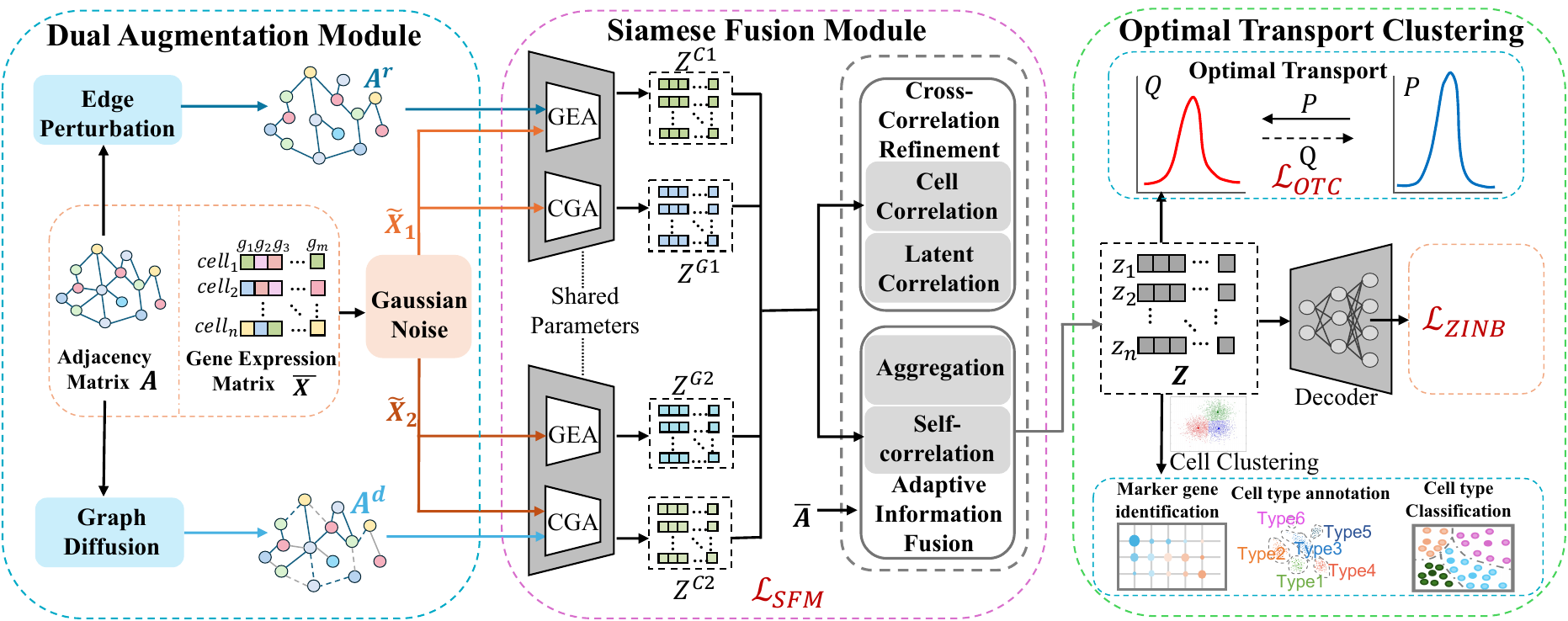}
    \caption{Overview architecture of~\methodname. It contains three components: (i) a data-augmented module, (ii) a Siamese fusion module, and (iii) an optimal transport clustering strategy for self-supervision learning.}
    \label{fig:framework_scSiamesesClu}
\end{figure*}
\section{Methodology}
This section introduces the proposed~\methodname, including the highlighted dual augmentation module, siamese fusion module, optimal transport clustering.
More details are provided in the supplementary material, e.g., gene expression autoencoders (GEAs) and cell graph autoencoders (CGAs). 

\subsection{Problem Formulation}
Given a single-cell expression matrix $\mathbf{X} \in \mathbb{R}^{N\times D}$, where $x_{ij}\left(1\leq i\leq N, 1\leq j\leq D\right)$ shows the expression of the $j$-th gene in the $i$-th cell. 
We construct an undirected graph $\mathcal{G_C}=(\mathcal{V}, \mathcal{E})$ using KNN based on the gene expression matrix, where $\mathcal{V}=\{v_1, v_2, \cdots, v_N \}$ is the set of $N$ nodes, $\mathcal{E}$  is the set of edges. 
Each node corresponds to a single cell, and edges indicate similarity between cells, quantified with the Pearson Correlation Coefficient~\cite{ding2022data,lu2022pearnet}. 
The KNN graph is defined with the preprocessed data matrix $\mathbf{\bar{X}}$, as specified in Section 4.1, and adjacency matrix $\mathbf{A} = (a_{ij})_{N \times N}$, where $a_{ij}=1$ if $v_i$ and $v_j$ are connected, and $a_{ij}=0$, otherwise.
The associated degree matrix is $\mathbf{D} = diag(d_1, d_2, \cdots, d_N ) \in \mathbb{R}^ {N \times N}$, where $d_i=\sum_{(v_i,v_j ) \in \mathcal{E}}{a_{ij}}$ represents the degree of node $v_i$. 
To account for self-loops and normalize the adjacency matrix, we compute $\widetilde{\mathbf{A}} \in \mathbb{R}^{N \times N}$ as $\mathbf{D}^{-1}(\mathbf{A}+\mathbf{I})$, where $\mathbf{I} \in \mathbb{R}^{N \times N}$ is the identity matrix. 
Our goal is to partition the graph $\mathcal{G_C}$ into $\mathbf{C}$ categories by assigning each node $v_i \in \mathcal{V}$ to one of the $\mathbf{C}$ categories based on $\mathcal{E}$ and gene expression features. 

\subsection{Framework Overview}
We propose\textbf{~\methodname}, an enhanced graph autoencoder-based siamese clustering framework for interpreting scRNA-seq data. 
We aim to improve the performance of scRNA-seq clustering and related downstream tasks by learning more accurate and distinctive cell embeddings. 
As Fig.~\ref{fig:framework_scSiamesesClu} shows, our framework consists of three components:
(i) a dual augmentation module that enhances the gene expression matrix and the cell graph relationships; 
(ii) a siamese fusion module to map the augmented data into a lower-dimensional latent space, generating cell embeddings; 
and (iii) an optimal transport clustering strategy for self-supervision learning, which refines the latent representations and generates clustering assignments. 

\subsection{Dual Augmentation Module}
To enhance robustness against noise and generalization across diverse datasets, biologically plausible augmentations are introduced at both gene and cell levels. 
These augmentations capture the variation in scRNA-seq, helping the model extract biologically meaningful representations. 

\noindent\textbf{Gene Expression Augmentation.} 
To enhance robustness at the gene level, biologically plausible augmentations are introduced by adding controlled noise to the gene expression profiles. 
Specifically, Gaussian noise is applied to simulate natural variability in gene expression. 
Let $\odot$ denote the Hadamard product~\cite{horn1990hadamard}, and $\mathbf{N} \in \mathbb{R}^{N \times D}$ represent a random noise matrix drawn from a Gaussian distribution $\mathcal{N}(1,0.1)$. 
The augmented gene expression matrix $\widetilde{\mathbf{X}}$ is computed as:  
\begin{equation}
\widetilde{\mathbf{X}}=\bar{\mathbf{X}} \odot \mathbf{N},
\label{equ:Gene Perturbation}
\end{equation}
as illustrated in Fig.~\ref{fig:framework_scSiamesesClu}, two augmented gene expression matrices, $\widetilde{\mathbf{X}}_1$ and $\widetilde{\mathbf{X}}_2$, were generated using Gaussian distributions with distinct parameters.

\noindent\textbf{Cell Graph Augmentation.} 
Cell graph augmentation aims to improve model robustness and enhance representation learning by introducing perturbations to the cell graph, specifically through edge perturbation and graph diffusion \cite{wang2024comprehensive}. 
These two strategies provide distinct, yet complementary, perspectives on the cell graph, enabling the model to capture diverse interactions between cells. 

\noindent(i) \noindent\textit{Edge perturbation.}  
We employ an edge removal strategy to refine the cell graph while preserving the most meaningful and biologically relevant relationships. 
Removing weaker edges reduces graph noise and improves model robustness. 
In contrast, edge addition is avoided to prevent spurious connections that may distort the graph and compromise its biological validity. 
Specifically, a mask matrix $\mathbf{M} \in \mathbb{R}^{N \times N}$ is constructed based on the  pairwise cosine similarity matrix computed in the latent space. 
The 10\% of edges with the lowest similarity values are identified and removed using this mask. 
Finally, the adjacency matrix $\mathbf{A}^r \in \mathbb{R}^{N \times N}$ is normalized for proper scaling, calculated as:
\begin{equation}
\mathbf{A}^r=\mathbf{D}^{-\frac{1}{2}}((\mathbf{A} \odot \mathbf{M})+\mathbf{I}) \mathbf{D}^{-\frac{1}{2}}.
\end{equation}
\noindent (ii) \noindent\textit{Graph diffusion.} 
We use a graph diffusion strategy to refine the cell graph by enhancing meaningful relationships between cells~\cite{hassani2020contrastive}. 
Specifically, the normalized adjacency matrix is transformed into a diffusion adjacency matrix using the Personalized PageRank (PPR) algorithm~\cite{page1998pagerank}. 
The teleport probability controls information propagation across the graph, helping to emphasize biologically meaningful relationships. 
The diffusion adjacency matrix $ \mathbf{A}^d$ would be computed as:
\begin{equation}
\mathbf{A}^d=\alpha\left(\mathbf{I}-(1-\alpha)\left(\mathbf{D}^{-\frac{1}{2}}(\mathbf{A}+\mathbf{I}) \mathbf{D}^{-\frac{1}{2}}\right)\right)^{-1},
\label{equ:Cell Perturbation}
\end{equation}
Thus, we define two distinct adjacency matrices, $\mathbf{A}^r$ and $\mathbf{A}^d$, from edge perturbation and graph diffusion, respectively.

\subsection{Siamese Fusion Module}
We propose the Siamese Fusion Module (\textbf{SFM}), a novel framework that integrates cross-correlation refinement and adaptive information fusion to learn discriminative and robust sample representations and avoid the over-smoothing issue of GNN-based methods~\cite{ning2022graph}. 
By minimizing the deviation of cross-correlation matrices from the ideal identity matrix across cells and genes and adaptive information fusion, SFM effectively reduces redundancy while preserving critical information~\cite{zbontar2021barlow,tu2021deep,liu2022deep}. 
This enhances the cell representations and avoids representation collapse, enabling more accurate cell population separation and improving clustering performance.

\noindent\textbf{Cross-Correlation Refinement.} 
The cross-correlation refinement process is designed to effectively encode and integrate the augmented data from genes and cells, ensuring the information from different augmentations is unified. 
We construct two gene expression autoencoders (GEAs) to process the augmented gene expression matrices $\widetilde{\mathbf{X}}_1$ and $\widetilde{\mathbf{X}}_2$, generating gene expression embeddings $\mathbf{Z}^{G1} \in \mathbb{R}^{N\times d}$ and $\mathbf{Z}^{G2} \in \mathbb{R}^{N\times d}$, where $d$ represents the latent embedding dimensionality. 
Similarly, we construct two cell graph autoencoders (CGAs) to process the augmented cell graph matrices ${\widetilde{\mathbf{X}}_1, \mathbf{A}^r}$ and ${\widetilde{\mathbf{X}}_2, \mathbf{A}^d}$, producing cell graph embeddings $\mathbf{Z}^{C1} \in \mathbb{R}^{N\times d}$ and $\mathbf{Z}^{C2} \in \mathbb{R}^{N\times d}$.
 
\noindent \emph{Cell Correlation Refinement.} 
Cell correlation refinement (CCR) optimizes cell relationships across different augmented views by aligning corresponding embeddings and reducing redundant correlations. 
We can calculate the refined cell correlation matrix by
$\mathbf{R}_{i j}^{C1}=\frac{\left({\mathbf{Z}}_i^{G1}\right)\left({\mathbf{Z}}_j^{G2}\right)^{\mathrm{T}}}{\left\|{\mathbf{Z}}_i^{G1}\right\|\left\|{\mathbf{Z}}_j^{G2}\right\|}, 
\mathbf{R}_{i j}^{C2}= \frac{\left({\mathbf{Z}}_i^{C1}\right)\left({\mathbf{Z}}_j^{C2}\right)^{\mathrm{T}}}{\left\|{\mathbf{Z}}_i^{C1}\right\|\left\|{\mathbf{Z}}_j^{C2}\right\|}, 
\forall i, j \in[1, N]$. 

To ensure the cell correlation matrix $\mathbf{R}^{C1},\mathbf{R}^{C2}$ aligns with an identity matrix ${\mathbf{I^{C}}} \in \mathbb{R}^{N \times N}$, we minimize loss: 

\begin{equation}
\small
\mathcal{L}_{Cor1} =\frac{1}{N^2} \sum\left(\mathbf{R}^{C1}-{\mathbf{I}}^C\right)^2  + \frac{1}{N^2} \sum\left(\mathbf{R}^{C2}-{\mathbf{I}}^C\right)^2 
\label{equ:loss_CCR}
\end{equation}
By ensuring the diagonal elements of $\mathbf{R}^{C1},\mathbf{R}^{C2}$ are equal to 1 and the off-diagonal elements are equal to 0, we can guarantee the alignment of embeddings for each view and minimize the consistency of embeddings for different cells across different views. 
This helps~\methodname~reduce redundant information and learn more discriminative representations. 

\noindent \emph{Latent Correlation Refinement.} 
Similarly, latent correlation refinement (LCR) optimizes relationships between latent embeddings across different augmented views by aligning corresponding embeddings. 
First, we project gene expression embeddings $\mathbf{Z}^{G1}$ and $\mathbf{Z}^{G2}$ into $\widetilde{\mathbf{Z}}^{G1}$ and $\widetilde{\mathbf{Z}}^{G2} \in \mathbb{R}^{d\times K}$ with using a readout function $\mathcal{R}(\cdot): \mathbb{R}^{d \times N} \rightarrow \mathbb{R}^{d \times K}$ , formulated as $\widetilde{\mathbf{Z}}^{G1}=\mathcal{R}\left(\left(\mathbf{Z}^{G1}\right)^{\mathrm{T}}\right)$, $\widetilde{\mathbf{Z}}^{G2}=\mathcal{R}\left(\left(\mathbf{Z}^{G2}\right)^{\mathrm{T}}\right)$. 
Likewise, we can project cell graph embeddings $\mathbf{Z}^{C1}$ and $\mathbf{Z}^{C2}$ into $\widetilde{\mathbf{Z}}^{C1}=\mathcal{R}\left(\left(\mathbf{Z}^{C1}\right)^{\mathrm{T}}\right)$ and $\widetilde{\mathbf{Z}}^{C2}=\mathcal{R}\left(\left(\mathbf{Z}^{C2}\right)^{\mathrm{T}}\right)$. 

Then we can calculate the refine latent correlation matrix by 
$\mathbf{R}_{i j}^{L1}=\frac{\left(\widetilde{\mathbf{Z}}_i^{G1}\right)\left(\widetilde{\mathbf{Z}}_j^{G2}\right)^{\mathrm{T}}}{\left\|\widetilde{\mathbf{Z}}_i^{G1}\right\|\left\|\widetilde{\mathbf{Z}}_j^{G2}\right\|}, 
\mathbf{R}_{i j}^{L2}= \frac{\left(\widetilde{\mathbf{Z}}_i^{C1}\right)\left(\widetilde{\mathbf{Z}}_j^{C2}\right)^{\mathrm{T}}}{\left\|\widetilde{\mathbf{Z}}_i^{C1}\right\|\left\|\widetilde{\mathbf{Z}}_j^{C2}\right\|}, 
\forall i, j \in[1, d]$, 
which denotes the correlation between the embeddings learned from enhanced gene expression matrix. 

To ensure the latent correlation matrix $\mathbf{R}^{L1},\mathbf{R}^{L2}$ aligns with an identity matrix ${\mathbf{I}^{L}} \in \mathbb{R}^{d \times d}$, we minimize the loss: 
\begin{equation}
\small
\mathcal{L}_{Cor2} =\frac{1}{d^2} \sum\left(\mathbf{R}^{L1}-{\mathbf{I}}^L\right)^2  + \frac{1}{d^2} \sum\left(\mathbf{R}^{L2}-{\mathbf{I}}^L\right)^2 
\label{equ:loss_LCR}
\end{equation}

\noindent\textbf{Adaptive Information Fusion.} 
To enhance clustering performance, we propose an adaptive information fusion mechanism that integrates cell relationships through embedding aggregation, self-correlation learning, and dynamic recombination. 
We first aggregate the embeddings achieved with GEA and CGA with a linear combination operation:
\begin{equation}
    \mathbf{Z}^{A} = \left(\mathbf{Z}^{G1} + \mathbf{Z}^{G2} + \mathbf{Z}^{C1} + \mathbf{Z}^{C2}\right)/4,
\label{equ:Z_i}
\end{equation}

Then, we apply a graph convolution-like operation (i.e., message passing) to process the combined information, enhancing the initial fused embedding $\mathbf{Z}^{A} \in \mathbb{R}^{N \times d}$. 
Specifically, we enhance $\mathbf{Z}^{A}$ by propagating cell graph $\widetilde{\mathbf{A}}$, i.e., 
$\mathbf{Z}^{E}=\widetilde{\mathbf{A}}\mathbf{Z}^{A}$. 
To capture the relationships among cell graphs in the enhanced embedding space, we compute a normalized self-correlation matrix $\mathbf{R^{S}} \in \mathbb{R}^{N \times N}$. 
Each element of $\mathbf{R^{S}}$ is defined as $\mathbf{R^{S}}_{i j}=\frac
{exp{\left(\left(\mathbf{Z}^E {\left(\mathbf{Z}^{E}\right)}^{\mathbf{T}}\right)_{i j}\right)}}
{\sum_{k=1}^N exp{\left(\left(\mathbf{Z}^E {\left(\mathbf{Z}^{E}\right)}^{\mathbf{T}}\right)_{i k}\right)}}$. 
The self-correlation matrix $\mathbf{R^S}$ encodes the relative similarity between cells, providing a mechanism to model relationships in the embedding space. 
Thus, we can recombine the embedding matrix $\mathbf{Z}^{E}$ to dynamically adjust the embedding representations. 
Specifically, the recombined embedding matrix $\mathbf{Z}^{R}$ is computed as $\mathbf{Z}^{R}=\mathbf{R^{S}}\mathbf{Z}^{E}$. 
Each cell integrates information from others through learned relationships, enhancing embeddings for clustering. 

To preserve the initial fused embedding information while incorporating the recombined embeddings, we adopt a skip connection to fuse $\mathbf{Z}^{E}$ and $\mathbf{Z}^{R}$. 
The final clustering-oriented latent embeddings $\mathbf{Z} \in \mathbb{R}^{N \times d}$ are computed as:
\begin{equation}
    \mathbf{Z}=\mathbf{Z}^{E} + \beta\mathbf{Z}^{R},
\label{equ:embd_z_middle}
\end{equation}
where $\beta$ is a learnable scale parameter, initialized to 0 and updated during training. 
This fusion mechanism effectively filters out redundant information and preserves discriminative features in the latent space, enabling the network to learn robust and meaningful representations that enhance clustering performance while avoiding representation collapse.

\noindent\textbf{Propagated Regularization.} 
The reconstruction loss, $\mathcal{L}_{REC}$ minimizes the joint mean square error (MSE) of reconstruction loss of gene expression and cell graph matrix. 
To further enhance representation quality, we introduce a propagation regularization. 
Drawing inspiration from~\cite{liu2022deep}, formulated as follows:
\begin{equation}
\mathcal{L}_R=JSD(\mathbf{Z}, \widetilde{\mathbf{A}} \mathbf{Z}),
\label{equ:loss_r}
\end{equation}
where $JSD\left(\cdot\right)$ refers the Jensen-Shannon divergence~\cite{fuglede2004jensen}. Utilizing Eq.~\ref{equ:loss_r}, the network can capture long-range dependencies even with a shallow network architecture, thereby reducing over-smoothing as the propagated information deepens within the framework. 
Finally, the overall objective of the SFM module can be computed by:
\begin{equation}
\mathcal{L}_{SFM}=\mathcal{L}_{Cor1}+\mathcal{L}_{Cor2}+\mathcal{L}_{REC}+\gamma\mathcal{L}_R,
\label{equ:loss_dicr}
\end{equation}
where $\gamma$ is a trade-off parameter balancing the contribution of the propagation regularization. 

\subsection{Optimal Transport Clustering} 
We utilize a self-supervised method to perform unsupervised clustering of scRNA-seq data~\cite{xie2016unsupervised}, which learn directly from the data without requiring labeled inputs. 
Specifically, we employ Student's t-distribution \cite{van2008visualizing} as a kernel function to measure the similarity $q_{ij}$ between each cell embedding $h_{i}$ and the cluster center $c_{j}$. 
By assigning higher weights to closer points, this method effectively captures the non-linear relationships in scRNA-seq data. 
The assignment distribution is represented as a matrix $Q=\left[q_{i j}\right]$, where each element indicates the probability or similarity between sample $i$ and cluster center $j$. 
To further refine clustering, we compute a target distribution matrix $P=\left[p_{i j}\right]$, which sharpens the soft assignments in $Q$ by emphasizing high-confidence samples. 
This sharpening process improves cluster separation and ensures well-defined clusters.
The clustering process is optimized by minimizing the divergence between $Q$ and 
$P$, iteratively refining results to group similar cells and separate dissimilar ones. 
In the target distribution $P$, each assignment in $Q$ is squared and normalized to enhance assignment confidence~\cite{bo2020structural}. 

To avoid degenerate solutions that would allocate all data points to a single label, we establish constraints that synchronize the label distribution with the mixing proportions. 
This ensures that each cell contributes equally to the loss calculation, improving clustering accuracy and preserving a balanced effect during learning process.  
We construct the target probability matrix $P$ using the following optimal transport strategy to ensure this alignment and enhance the robustness of our clustering outcomes: 
\begin{equation}
\small
\centering
\begin{aligned}
        \min_{P}  & -P*(\text{log}Q) \\
        \mbox{s.t.} & P\in\mathbb{R}_{+}^{N\times C}, 
         P\ \bm{1}_C = \bm{1}_N \ \text{ and } \ P^{T}\bm{1}_N = N\bm{\pi}.
\end{aligned}
\label{equ:first_ot_optimization function}
\end{equation}
In this context, we regard the target distribution $P$ as the transport plan matrix derived from optimal transport theory, while $- \text{log}Q$ serves as the cost matrix. 
We impose the constraint $P^{T}\bm{1}_N = N\bm{\pi}$, where $\pi$ indicates the proportion of cells assigned to each cluster, estimated from intermediate clustering results. 
Thus, we can ensure that the resulting cluster distribution aligns with the defined mixing proportions. 
Given the substantial computational expense associated with direct optimization techniques, we utilize the Sinkhorn distance~\cite{sinkhorn1967diagonal} to facilitate quicker optimization through an entropic constraint. 
The optimization problem, which includes a Lagrange multiplier for the entropy constraint is formulated as: 
\begin{equation}
\begin{aligned}
        \min_{P} & -P*(\text{log}Q)-\frac{1}{\lambda}\text{H}(P) \\
        \mbox{s.t.} & P\in\mathbb{R}_{+}^{N\times C}, 
         P\ \bm{1}_C = \bm{1}_N \ \text{ and } \ P^{T}\bm{1}_N = N\bm{\pi},
\end{aligned}
\label{equ:opt_optimization function}
\end{equation}
where $\bm{H}$ is the entropy function  measuring uncertainty in the distribution, and $\lambda$ is the smoothness parameter that maintains cluster balance.
The transport plan $P$ is guaranteed to exist and be unique, with the solution  efficiently obtained through Sinkhorn's method~\cite{sinkhorn1967diagonal} as follows: 
\begin{equation}
\hat{P}^{(t)}=\operatorname{diag}\left(\boldsymbol{u}^{(t)}\right) Q^\lambda \operatorname{diag}\left(\boldsymbol{v}^{(t)}\right)
\end{equation}
at each iteration $t$, $\bm{u}^{(t)}$ is updated as $\bm{1}_N / (Q^{\lambda}\bm{v}^{(t-1)})$ and $\bm{v}$ is calculated as $N_\pi / (Q^{\lambda}\bm{u}^{(t)})$. 
We begin with the initial value set as $\bm{v}^{(0)}=\bm{1}_N$. 
Upon convergence, the optimal transport plan matrix $\hat{P}$ is obtained. 

During training, $\hat{P}$ is fixed and $Q$ is aligned with $\hat{P}$. 
This alignment is crucial for evaluating model performance, allowing us to define the clustering loss function as follows: 
\begin{equation}
\mathcal{L}_{OTC}=\text{KL}(\hat{P} \| Q)=\sum_i \sum_j \hat{p}_{i j} \log \frac{\hat{p}_{i j}}{q_{i j}}
\label{equ:loss_clu}
\end{equation}

\subsection{Joint Optimization}
The Zero-Inflated Negative Binomial (ZINB) loss $\mathcal{L}_{ZINB}$ is commonly used to handle the sparsity and overdispersion of scRNA-seq data by modeling excess zeros and variability, 
ensuring accurate data reconstruction~\cite{eraslan2019single}. 
Thus, in the proposed method, the overall optimization objective consists of three parts: the SFM loss, the OTC loss, and the ZINB loss, which are formulated as:
\begin{equation}
\mathcal{L}=\mathcal{L}_{SFM}+\rho\mathcal{L}_{ZINB}+\sigma\mathcal{L}_{OTC},
\label{equ:loss_all}
\end{equation}
where the hyperparameters $\rho$ and $\sigma$ balance the components, facilitating effective embedding learning and clustering. 

\begin{table*}[!th]
    \centering
    \small
    \resizebox{\textwidth}{!}
    { 
    \renewcommand\arraystretch{1.2}
    \begin{tabular}{c|c|cc|cccc|ccc|c}
        \toprule
        \textbf{\large Dataset} & \textbf{\large Metric} & \textbf{\large \makecell{\large pcaReduce}} & \textbf{\large \makecell{SUSSC}} & \textbf{\large \makecell{DEC}} & \textbf{\large \makecell{scDeepCluster}} & \textbf{\large scDCC} & \textbf{\large scNAME} & \textbf{\large scDSC} & \textbf{\large scGNN} & \textbf{\large \makecell{scCDCG}}& \textbf{\large \makecell{OURS}}\\
        \midrule
        
        \multirow{3}*{\textbf{\large \makecell{Shekhar mouse retina cells}}}
        & \large ACC & \large 21.27 \small$\pm$ 0.14  & \large 25.04  \small$\pm$ 0.73 & \large 24.55 \small$\pm$ 2.29 & \large 47.26 \small$\pm$ 3.34 &\large 74.14 \small$\pm$ 2.31 & \large \underline{79.84} \small$\pm$ 2.77 & \large 67.54 \small$\pm$ 4.08 &\large 78.56 \small$\pm$ 3.95 & \large 76.04 \small$\pm$ 1.85 &\large \textbf{89.16} \small$\pm$ 3.52\\
        & \large NMI & \large 25.54 \small$\pm$ 0.05  & \large 43.35  \small$\pm$ 0.36 & \large 40.32 \small$\pm$ 3.50 & \large 80.45 \small$\pm$ 1.28 &\large \underline{81.20} \small$\pm$ 0.68 & \large 75.19 \small$\pm$ 11.89 & \large 62.46 \small$\pm$ 14.53 &\large 79.89 \small$\pm$ 5.28 & \large 76.71 \small$\pm$ 1.47 &\large \textbf{82.72} \small$\pm$ 1.22\\
        & \large ARI & \large 10.01 \small$\pm$ 0.03  & \large 19.69  \small$\pm$ 0.52 & \large 17.96 \small$\pm$ 3.59 & \large 52.66 \small$\pm$ 5.77 &\large 62.44 \small$\pm$ 4.50 & \large \underline{81.66} \small$\pm$ 7.05 & \large 54.53 \small$\pm$ 8.34 &\large 77.58 \small$\pm$ 9.59 & \large 60.65 \small$\pm$ 6.47 &\large \textbf{88.93} \small$\pm$ 2.15\\
        \hline
        
        \multirow{3}*{\textbf{\large \makecell{Macosko mouse retina cells}}}
        & \large ACC & \large 16.40 \small$\pm$ 0.33 & \large 30.34 \small$\pm$ 0.40 & \large 33.32 \small$\pm$ 2.97 & \large 54.45 \small$\pm$ 3.03 &\large 63.31 \small$\pm$ 11.09 & \large \underline{78.97} \small$\pm$ 7.76 & \large 72.16 \small$\pm$ 0.77 & \large 62.80 \small$\pm$ 2.26 & \large 69.78 \small$\pm$ 0.70 &\large \textbf{87.13} \small$\pm$ 2.15\\
        & \large NMI & \large 23.35 \small$\pm$ 0.19  & \large 50.72 \small$\pm$ 0.13 & \large 52.20 \small$\pm$ 2.14 & \large 76.83 \small$\pm$ 2.09 &\large 79.13 \small$\pm$ 0.94 & \large \underline{82.64} \small$\pm$ 1.44 & \large 67.04 \small$\pm$ 0.49 & \large 68.25 \small$\pm$ 3.36 & \large 68.71 \small$\pm$ 2.52 &\large \textbf{86.62} \small$\pm$ 2.31\\
        & \large ARI & \large 11.44 \small$\pm$ 0.79 & \large 18.89 \small$\pm$ 0.53 & \large 27.55 \small$\pm$ 6.65 & \large 40.90 \small$\pm$ 1.44 &\large 52.38 \small$\pm$ 2.78 & \large \underline{84.72} \small$\pm$ 0.17 & \large 61.55 \small$\pm$ 0.82 & \large 74.01 \small$\pm$ 2.25 & \large 58.99 \small$\pm$ 2.69 &\large \textbf{90.92} \small$\pm$ 3.74\\
        \hline

        \multirow{3}*{\textbf{\large \makecell{QS mouse lung cells}}}
        & \large ACC & \large 33.70 \small$\pm$ 0.21 & \large 40.82 \small$ \pm$ 0.10 & \large 38.07 \small$\pm$ 1.40 & \large 47.16 \small$\pm$ 3.35 &\large 48.29 \small$\pm$ 5.51 & \large 69.84 \small$\pm$ 2.77 & \large 67.54 \small$\pm$ 4.08 & \large 69.46 \small$\pm$ 1.88 & \large \underline{70.58} \small$\pm$ 0.87 &\large \textbf{87.16} \small$\pm$ 1.52\\
        & \large NMI & \large 16.10 \small$\pm$ 0.54 & \large 43.08 \small$ \pm$ 0.01 & \large 37.27 \small$\pm$ 3.76 & \large 70.18 \small$\pm$ 2.24 &\large 50.40 \small$\pm$ 7.24 & \large 71.41 \small$\pm$ 5.40 & \large 69.48 \small$\pm$ 7.11 & \large 64.20 \small$\pm$ 2.10 & \large \underline{70.13} \small$\pm$ 1.92 &\large \textbf{86.07} \small$\pm$ 1.04\\
        & \large ARI & \large 11.68 \small$\pm$ 0.25 & \large 28.07 \small$ \pm$ 0.04 & \large 23.08 \small$\pm$ 3.75 & \large 41.37 \small$\pm$ 4.07 &\large 42.76 \small$\pm$ 10.05 & \large 45.50 \small$\pm$ 3.20 & \large 76.76 \small$\pm$ 7.24 & \large 75.90 \small$\pm$ 0.97 & \large \underline{80.67} \small$\pm$ 0.78 &\large \textbf{84.
        49} \small$\pm$ 3.44\\
        \hline
        
        \multirow{3}*{\textbf{\large \makecell{CITE-CMBC}}}
        & \large ACC & \large 28.15 \small$\pm$ 0.07 & \large 47.19 \small $\pm$ 0.02 & \large 41.88 \small$\pm$ 5.35 & \large 67.59 \small$\pm$ 4.40 &\large 69.04 \small$\pm$ 3.46 & \large 60.22 \small$\pm$ 5.87 & \large 68.79 \small$\pm$ 0.87 & \large 66.71 \small$\pm$ 1.65 & \large \underline{71.75} \small$\pm$ 1.8 &\large \textbf{74.70} \small$\pm$ 0.51\\
        & \large NMI & \large 28.28 \small$\pm$ 0.19 & \large 60.14 \small$ \pm$ 0.04 & \large 46.87 \small$\pm$ 9.23 & \large 73.16 \small$\pm$ 1.29 &\large 72.83 \small$\pm$ 1.31 & \large 48.98 \small$\pm$ 7.06 & \large 64.21 \small$\pm$ 3.25 & \large 61.70 \small$\pm$ 3.21 & \large \underline{68.57} \small$\pm$ 1.7 &\large \textbf{68.71} \small$\pm$ 0.75\\
        & \large ARI & \large 5.09 \small$\pm$ 0.04 & \large 32.84 \small $\pm$ 4.51 & \large 23.55 \small$\pm$ 10.25 & \large 51.21 \small$\pm$ 3.40 &\large 52.42 \small$\pm$ 2.60 & \large 66.06 \small$\pm$ 3.37 & \large 52.51 \small$\pm$ 2.15 & \large 66.25 \small$\pm$ 1.75 & \large \underline{61.06} \small$\pm$ 1.4 &\large \textbf{67.56} \small$\pm$ 1.17\\
        \hline

        \multirow{3}*{\textbf{\large \makecell{Human liver cells}}}
        & \large ACC & \large 35.04 \small$\pm$ 0.14 & \large 47.96 \small$\pm$ 0.06 & \large 46.32 \small$\pm$ 6.76 & \large 61.32 \small$\pm$ 8.61 &\large 72.90 \small$\pm$ 4.02 & \large 74.55 \small$\pm$ 5.30 & \large 70.90 \small$\pm$ 2.23 & \large 72.33 \small$\pm$ 3.4 & \large \underline{75.34} \small$\pm$ 1.7 & \large \textbf{88.33} \small$\pm$ 1.74\\
        & \large NMI & \large 32.43 \small$\pm$ 0.33 & \large 67.17 \small $\pm$ 0.00 & \large 52.24 \small$\pm$ 3.44 & \large 78.09 \small$\pm$ 2.50 & \large 78.81 \small$\pm$ 1.50 & \large 77.42 \small$\pm$ 12.00 & \large 71.63 \small$\pm$ 2.75 & \large 73.56 \small$\pm$ 2.7 & \large \underline{79.34} \small$\pm$ 2.6 & \large \textbf{88.82} \small$\pm$ 1.24\\
        & \large ARI & \large 10.24 \small$\pm$ 1.42 & \large 35.27 \small$\pm$ 0.00 & \large 31.04 \small$\pm$ 10.90 & \large 58.28 \small$\pm$ 10.61 & \large 70.47 \small$\pm$ 1.93 & \large 79.91 \small$\pm$ 2.93 & \large75.34 \small$\pm$ 3.56 & \large 76.01 \small$\pm$ 2.6 & \large \underline{81.26} \small$\pm$ 2.7 &\large \textbf{91.90} \small$\pm$ 0.48\\
        \hline

        \multirow{3}*{\textbf{\large \makecell{ Human kidney cells}}}
        & \large ACC & \large 41.38 \small$\pm$ 0.24 & \large 44.18 \small $\pm$ 0.22 & \large 38.91 \small$\pm$ 2.71 & \large 59.81 \small$\pm$ 4.69 & \large 67.89 \small$\pm$ 6.22 & \large 73.71 \small$\pm$ 3.90 & \large 75.32 \small$\pm$ 2.81 & \large 77.73 \small$\pm$ 1.41 & \large \underline{79.55} \small$\pm$ 0.29 &\large \textbf{86.08} \small$\pm$ 0.89\\
        & \large NMI & \large 24.31 \small$\pm$ 0.13 & \large 37.35 \small $\pm$ 0.02 & \large 28.78 \small$\pm$ 5.10 & \large 63.67 \small$\pm$ 3.70 & \large 66.53 \small$\pm$ 4.91 & \large 61.39 \small$\pm$ 4.70 & \large 69.41 \small$\pm$ 2.12 & \large \underline{73.79} \small$\pm$ 1.07 & \large 67.98 \small$\pm$ 2.62 &\large \textbf{80.86} \small$\pm$ 0.96\\
        & \large ARI & \large 24.56 \small$\pm$ 0.29 & \large 30.78 \small $\pm$ 0.02 & \large 19.97 \small$\pm$ 5.06 & \large 47.53 \small$\pm$ 5.44 & \large 50.37 \small$\pm$  7.22 & \large 68.39 \small$\pm$ 2.41 & \large 67.74 \small$\pm$ 4.52 & \large \underline{72.29} \small$\pm$ 1.59 & \large 64.91 \small$\pm$ 0.88 &\large \textbf{74.23} \small$\pm$ 2.07\\
        \hline

        \multirow{3}*{\textbf{\large \makecell{Human pancreas cells}}}
        & \large ACC & \large 51.17 \small$\pm$ 0.66 & \large 76.29 \small $\pm$ 0.02 & \large 72.22 \small$\pm$ 5.51 & \large 74.70 \small$\pm$ 2.69 &\large 86.65 \small$\pm$ 5.76 & \large 89.63 \small$\pm$ 5.83 & \large 79.42 \small$\pm$ 1.34 & \large 79.32 \small$\pm$ 3.96 & \large \underline{92.65} \small$\pm$ 1.9 &\large \textbf{94.95} \small$\pm$ 0.15\\
        & \large NMI & \large 57.71 \small$\pm$ 0.67 & \large 82.31 \small $\pm$ 0.13 & \large 76.29 \small$\pm$ 2.73 & \large 79.32 \small$\pm$ 0.32 & \large 83.81 \small$\pm$ 2.94 & \large 86.70 \small$\pm$ 7.79 & \large 75.89 \small$\pm$ 0.61 & \large 78.76 \small$\pm$ 5.60 & \large \underline{85.81} \small$\pm$ 1.0  &\large \textbf{86.19} \small$\pm$ 0.34\\
        & \large ARI & \large 37.00 \small$\pm$ 0.45 & \large 64.24 \small $\pm$ 0.01 & \large 61.76 \small$\pm$ 5.55 & \large 64.59 \small$\pm$ 2.49 & \large 79.83 \small$\pm$ 10.79 & \large 84.93 \small$\pm$ 3.13 & \large 75.42 \small$\pm$ 1.22 & \large 78.81 \small$\pm$ 3.17 & \large \underline{91.37} \small$\pm$ 1.21 &\large \textbf{91.59} \small$\pm$ 0.35\\
        \bottomrule
    \end{tabular} }
    \caption{Clustering performance across seven datasets (mean $\pm$ standard deviation), with best results in \textbf{bold} and runner-up results in \underline{underline}.} 
    \label{tab:expriment}
\end{table*}

\begin{figure*}[thbp]
\centering
\subfloat[scDeepCluster]{
\includegraphics[width=0.14\textwidth]{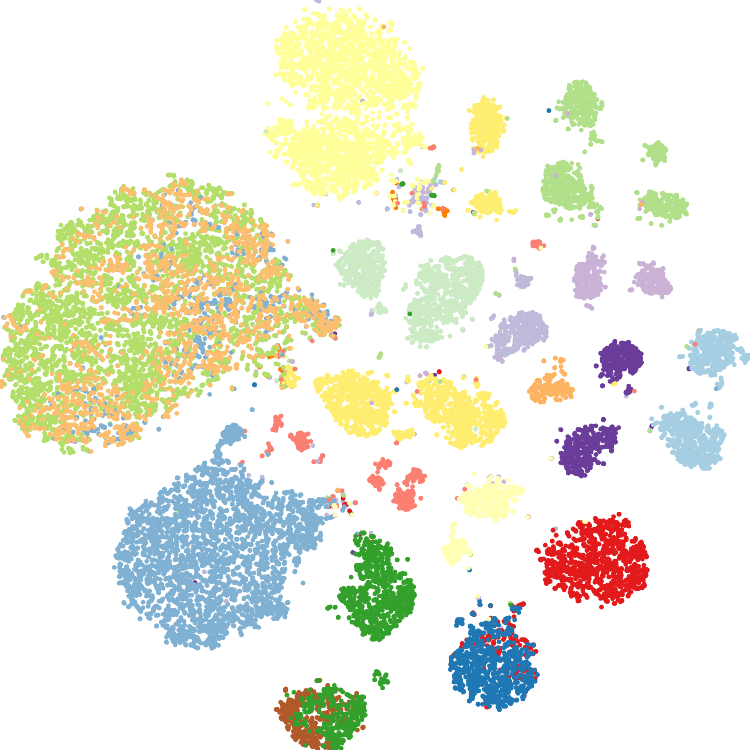}
}
\subfloat[scNAME]{
\includegraphics[width=0.14\textwidth]{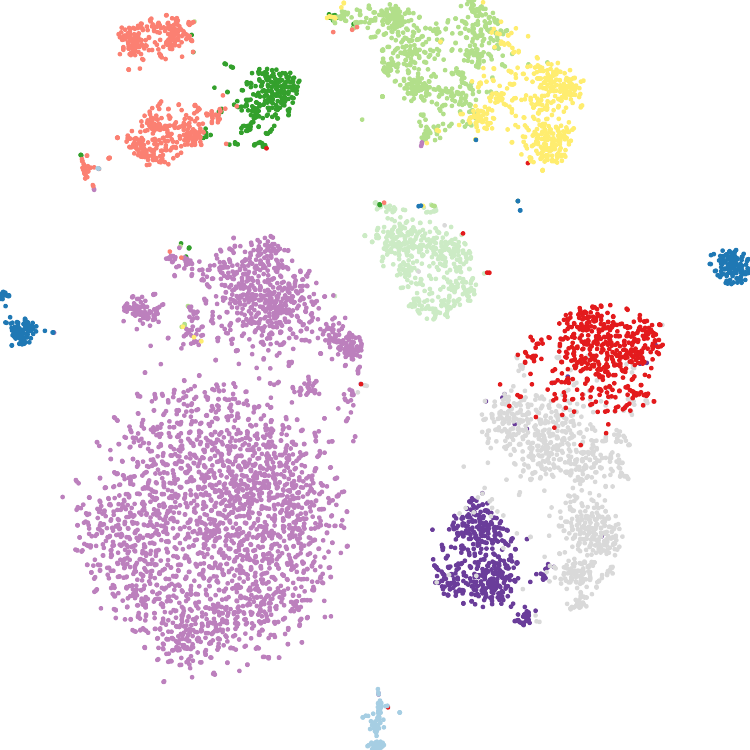}
}
\subfloat[scGNN]{
\includegraphics[width=0.14\textwidth]{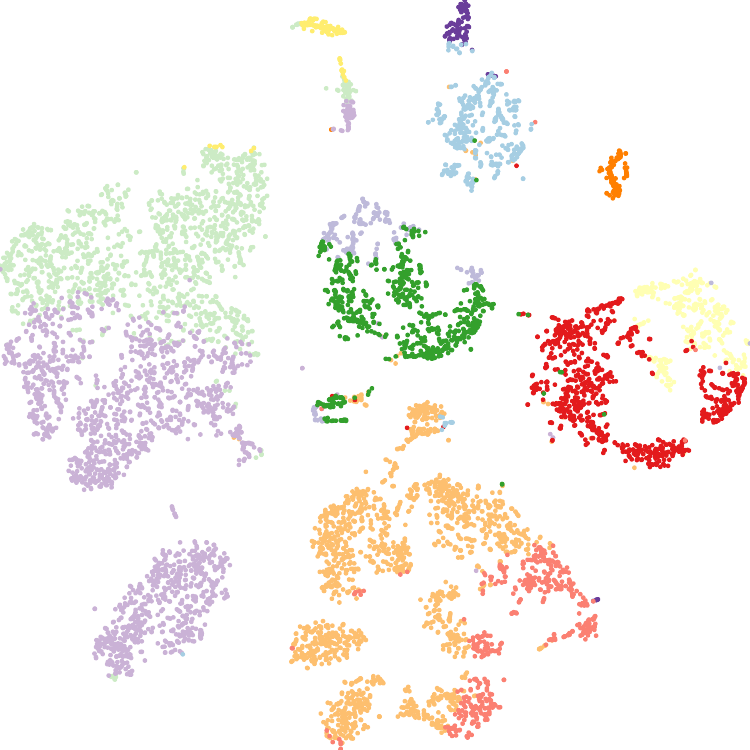}
}
\subfloat[scCDCG]{
\includegraphics[width=0.14\textwidth]{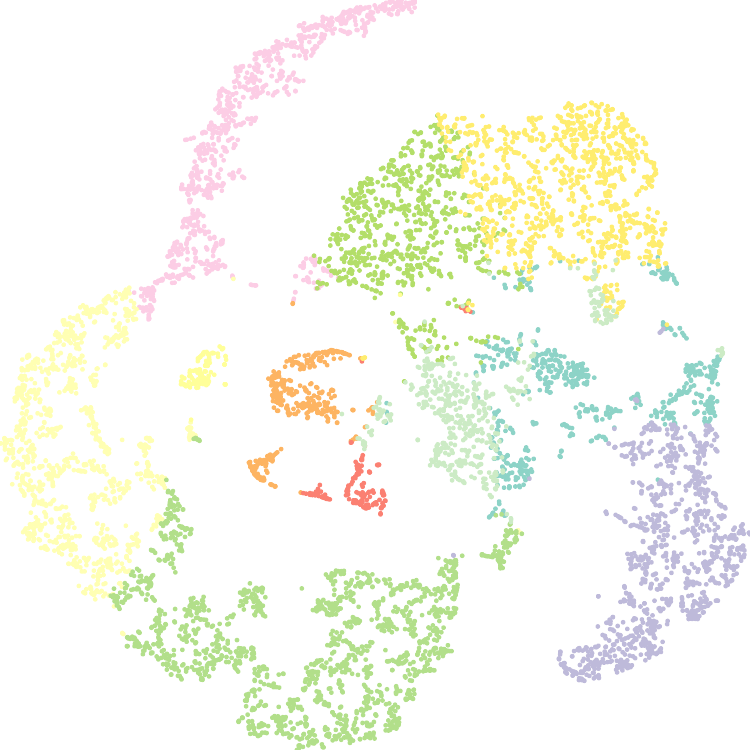}
}
\subfloat[scSiameseClu]{
\includegraphics[width=0.14\textwidth]{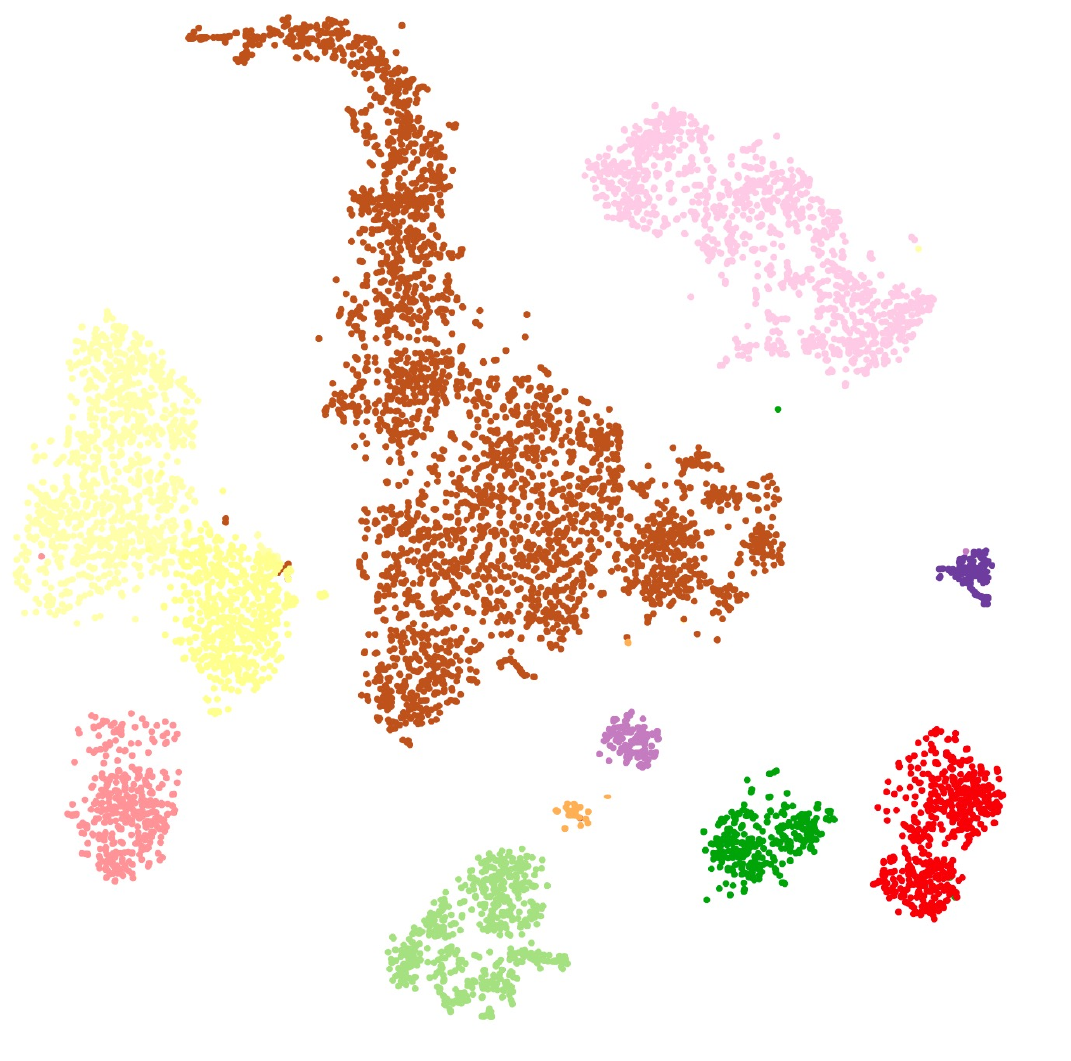}
}
\caption{Visualization of~\methodname~and four typical baselines on \emph{human liver cells} in 2D t-SNE projection. 
Each point represents a cell, while each color represents a predicted cell type.}
\label{fig:visualization}
\end{figure*}

\begin{figure}[!t]
\centering
\subfloat[Similarity distribution]{
\includegraphics[width=0.17\textwidth]{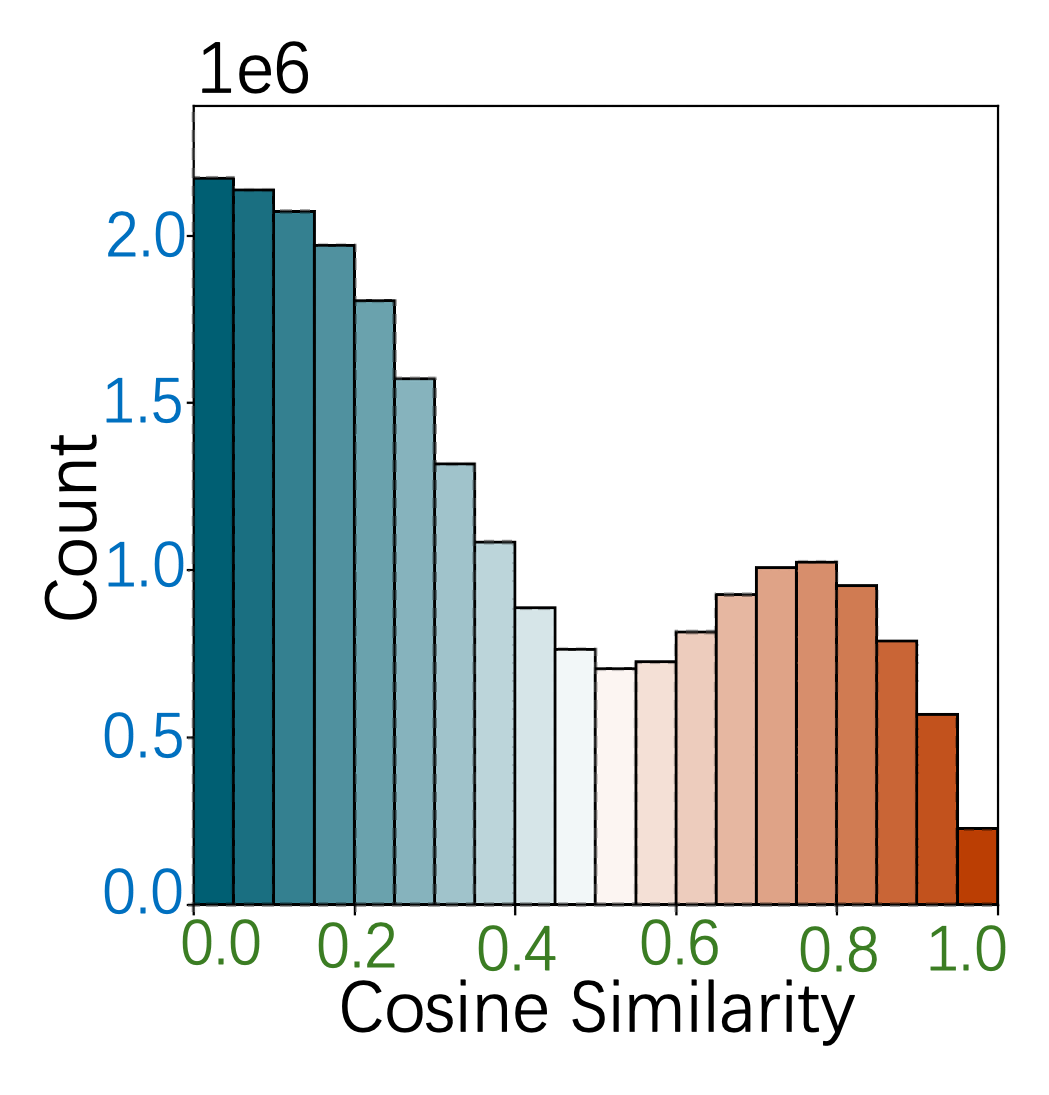}
}
\subfloat[Similarity heatmap]{
\includegraphics[width=0.205\textwidth]{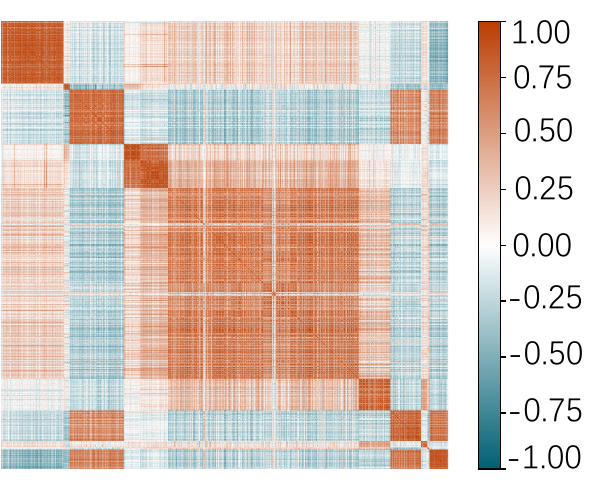}
}
\caption{Distribution plot and heat map of cell similarities in latent space learned by~\methodname~on the ~\emph{Human liver cells} dataset.}
\label{fig:exp_heatmap}
\end{figure}

\section{Experiments}  
In this section, we first validate our method through extensive experiments, demonstrating its superior performance in scRNA-seq clustering. 
We then show that the embeddings generated by our method effectively mitigate the embedding collapse issue in graph-based models. 
We also evaluate its utility in downstream tasks, like cell type annotation. 


\subsection{Experimental Setup}
\noindent\textbf{Dataset Preprocessing.} 
We evaluated~\methodname~on seven real scRNA-seq datasets. 
First, genes expressed in fewer than three cells were filtered out, 
followed by normalization, log-transformed (logTPM), and selection of highly variable genes based on predefined mean and dispersion thresholds. 
Finally, the preprocessed data were used as input.  


\noindent\textbf{Baseline Methods.} 
To verify the superior performance of the proposed approach, we compare it against nine competing SOTA clustering methods. 
Among them, pcaReduce~\cite{vzurauskiene2016pcareduce} and SUSSC~\cite{wang2021suscc} represent early traditional clustering techniques. 
Additionally, we consider four deep neural network-based clustering methods, DEC~\cite{xie2016unsupervised}, scDeepCluster~\cite{tian2019clustering}, scDCC~\cite{tian2021model}, and scNAME~\cite{wan2022scname}, which train an autoencoder to obtain representations followed by clustering. 
scDSC~\cite{gan2022deep}, scGNN~\cite{wang2021scgnn}, and scCDCG~\cite{xu2024sccdcg} are representative GNN-based clustering methods that incorporate both node attributes and structural information for latent representation.

\noindent\textbf{Implementation Details.}
We implemented ~\methodname~in Python 3.7 using PyTorch, with DCRN~\cite{liu2022deep} as the backbone. 
Each experiment was repeated 10 times, reporting the mean and variance. 
Detailed parameter settings are provided in the supplementary material. 

\noindent\textbf{Evaluation Metrics. } 
To illustrate the effectiveness of our approach, we evaluate clustering performance through three widely recognized metrics: Accuracy (ACC), Normalized Mutual Information (NMI)~\cite{strehl2002cluster}, and Adjusted Rand Index (ARI)~\cite{vinh2009information}.


\begin{figure*}[thbp]
\centering
\subfloat[scDeepCluster]{
\includegraphics[width=0.157\textwidth]{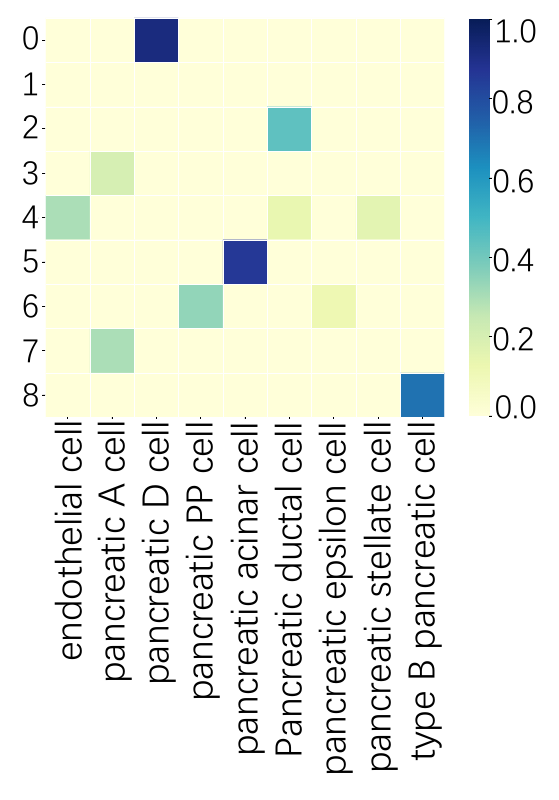}
}
\subfloat[scNAME]{
\includegraphics[width=0.157\textwidth]{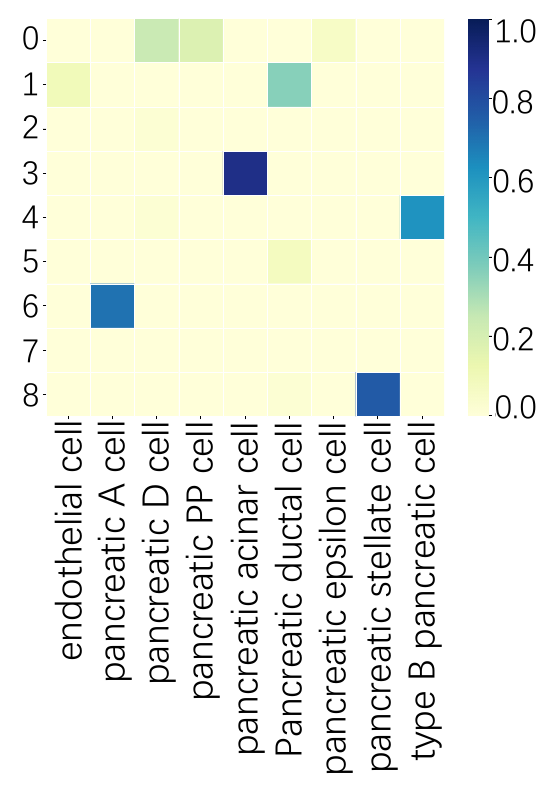}
}
\subfloat[scGNN]{
\includegraphics[width=0.157\textwidth]{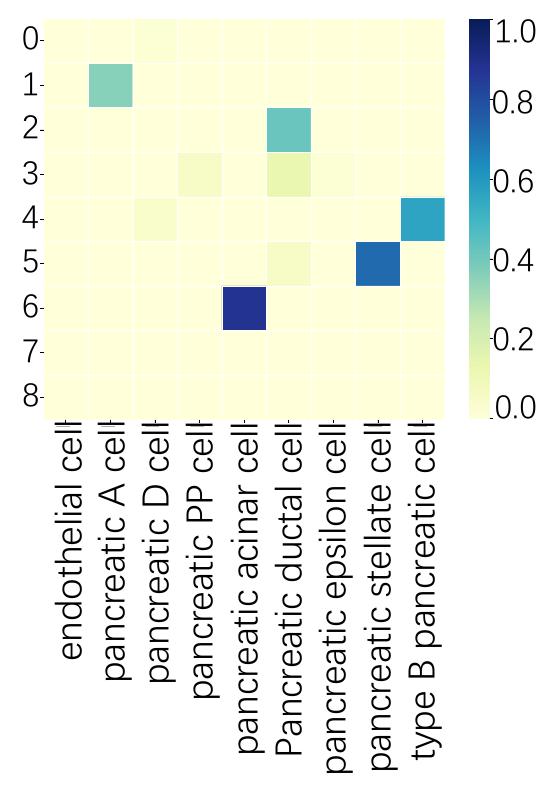}
}
\subfloat[scCDCG]{
\includegraphics[width=0.157\textwidth]{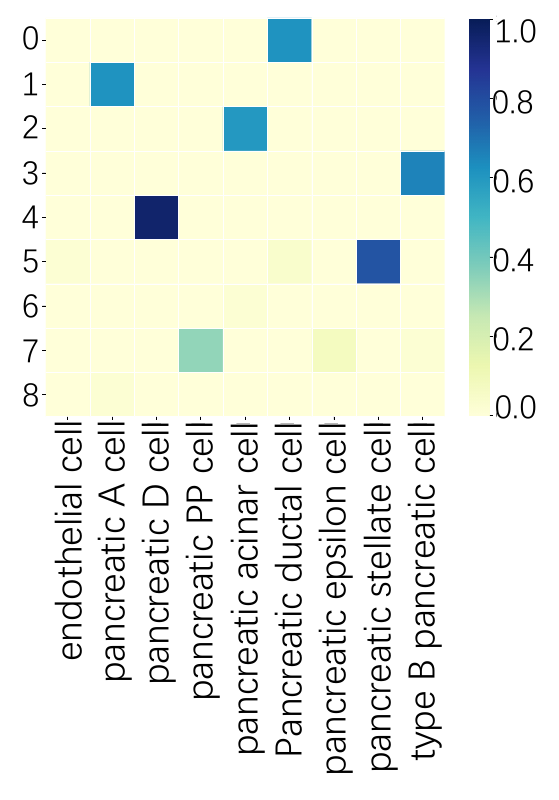}
}
\subfloat[\methodname]{
\includegraphics[width=0.157\textwidth]{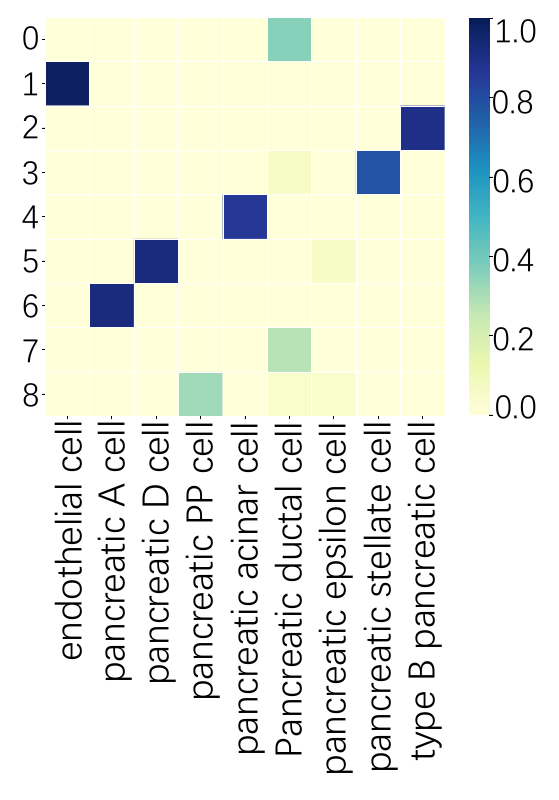}
}
\caption{Cell type annotation: overlap of top 50 DEGs in  clusters versus gold standard cell types (similarity = overlapping DEGs/50).}
\label{fig:CTA}
\end{figure*}

\begin{figure*}[t]
\centering
\subfloat[Gold Standard]{
\includegraphics[width=0.44\textwidth]{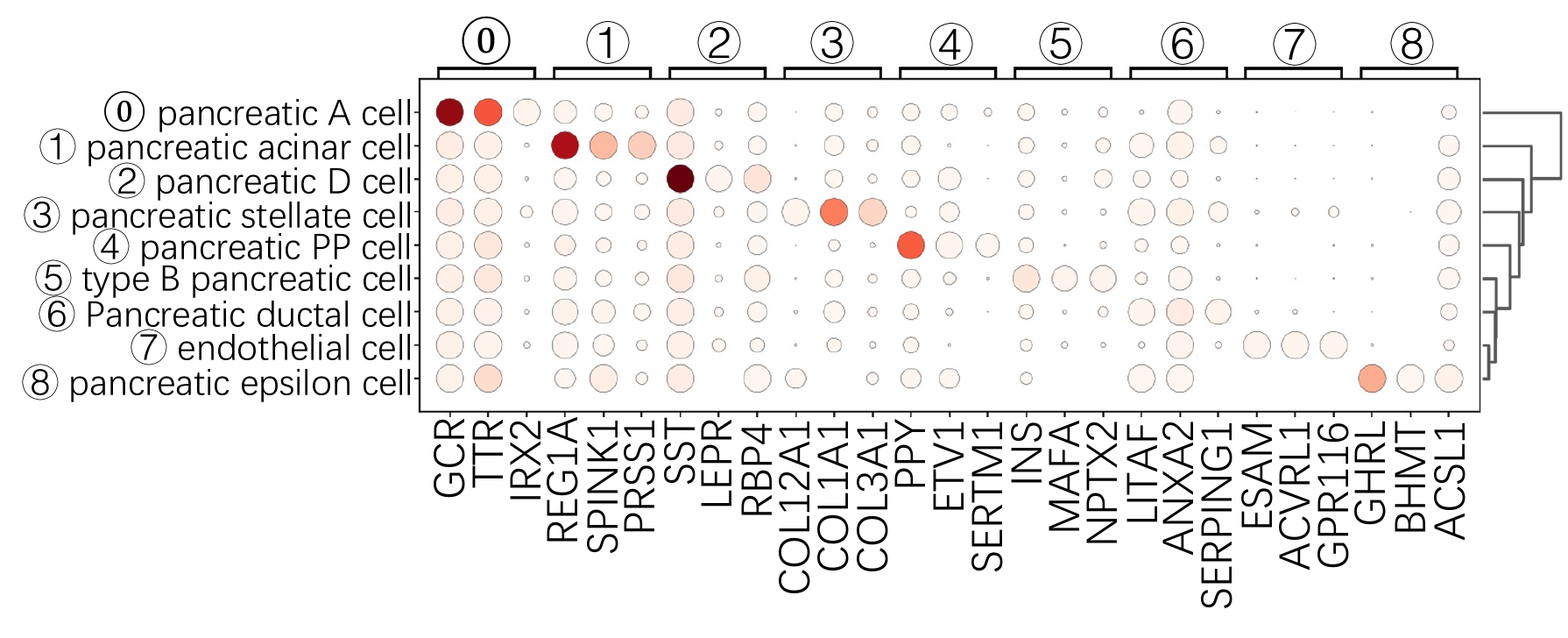}}
\subfloat[\methodname]{
\includegraphics[width=0.42\textwidth]{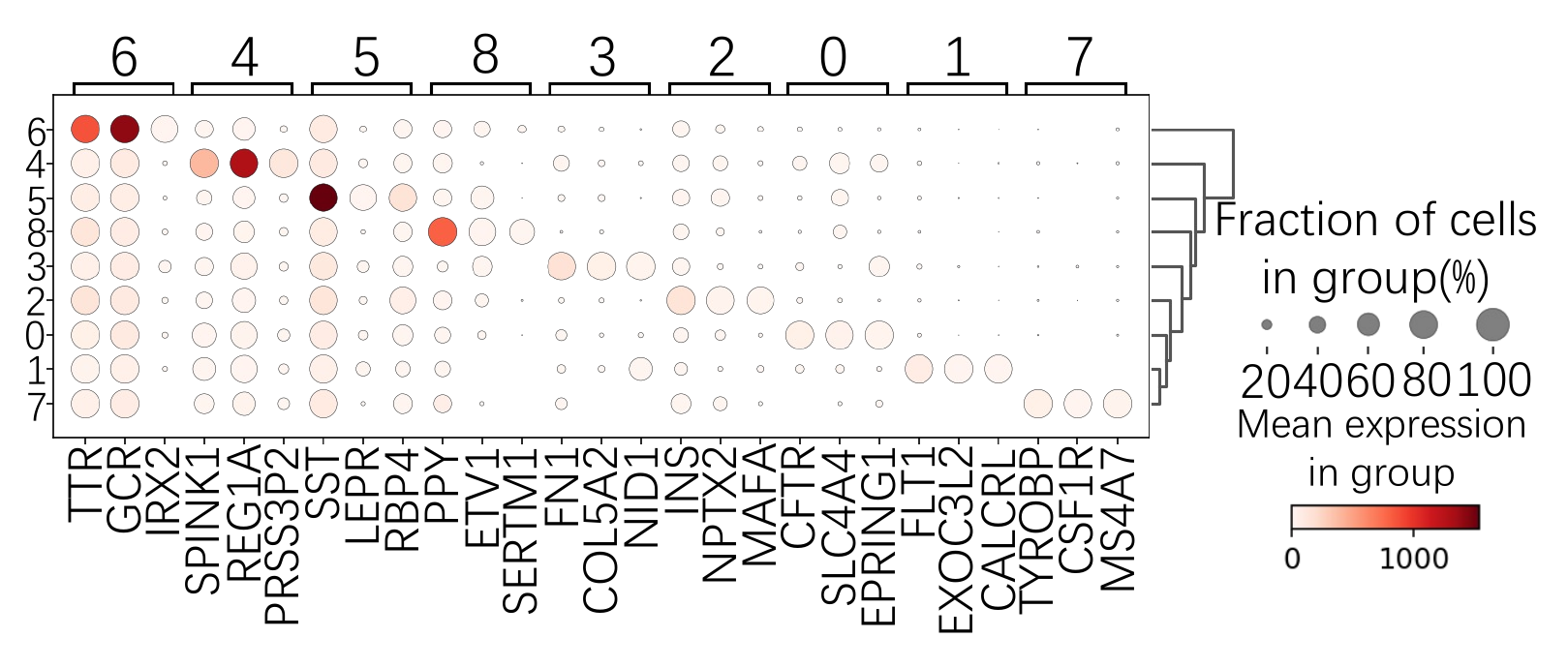}
}
\caption{Top 3 DEGs per cell type annotated by~\methodname~and gold standard.}
\label{fig:DEG}
\end{figure*}

\subsection{Overall Performance}
\noindent\textbf{Quantitative Analysis.} 
Tab.~\ref{tab:expriment} compares the clustering performance of our method with nine competitive baselines across seven benchmark datasets. 
Based on the results, we can observe that:  
1)~\methodname~significantly outperforms traditional single-cell clustering methods, demonstrating its superior ability to handle scRNA-seq data. 
2) Compared to deep learning-based methods, which rely solely on node information for clustering,~\methodname~achieves consistently better results, highlighting its ability to leverage both gene and cell information effectively. 
3) While GNN-based methods incorporate cell graph information, they suffer from redundant embeddings and representation collapse. 
In contrast,~\methodname~learns more discriminative representations of scRNA-seq data, thereby enhancing clustering performance. 
Overall,~\methodname~ outperforms the competing methods on three metrics across all datasets. 
On average,~\methodname~achieves a significant average improvement of 8.41\% in ACC, 5.89\% in NMI, 4.34\% in ARI than the second-best method.

\noindent\textbf{Qualitative Analysis.} 
To evaluate clustering performance, we used t-SNE to project the learned embeddings from different methods into a two-dimensional space on the~\emph{Human liver cells} dataset. 
As Fig.~\ref{fig:visualization} shows,~\methodname~produces well-separated clusters with clear boundaries, effectively distinguishing different cell types regardless of cluster size. 
In contrast, methods like scDeepCluster show dispersed clusters, failing to group similar cell types. 
We further analyze the embeddings from ~\methodname~with the cosine similarity matrices. 
As shown in Fig.~\ref{fig:exp_heatmap}, the bimodal distribution of cosine similarities (approaching a multivariate Gaussian) and the distinct heatmap patterns preserve meaningful cluster structures in the latent space. 
~\methodname~mitigates representation collapse and effectively captures latent cluster structures, unlike other deep learning and GNN-based methods (Fig\ref{fig:intro_heatmap}).

\subsection{Biological Analysis} 

\noindent\textbf{Cell Type Annotation.} 
To demonstrate that our clustering method enables more effective downstream analysis, we first identified differentially expressed genes (DEGs) and marker genes for each cluster using the "FindAllMarkers" function in Seurat~\cite{butler2018integrating}. 
Then, we compared the top 50 marker genes identified by~\methodname~and other methods to the gold standard on the~\emph{human pancreas cells} dataset, and our method accurately annotated clusters with cell types, achieving over 90\% similarity in marker genes for most clusters.
As Fig.~\ref{fig:CTA} shows, cluster 1 was identified as endothelial cell type, and cluster 2 as type B pancreatic cells, etc. 
Additionally, Fig.~\ref{fig:DEG} shows~\methodname~identifies marker genes for each cluster, e.g., cluster 6 is characterized as pancreatic A cell type with marker genes TTR, GCG, and IRX2. 
Our method effectively matched clusters with known cell types, providing reliable support for mechanistic studies.

\begin{table}[!t]
    \centering
    \resizebox{\linewidth}{!}{
    \begin{tabular}{l|ccccc|c}
         \toprule
         Metric & pcaReduce & scDCC & scDSC & scCDCG & scGNN  & \methodname \\
         \midrule
         Accuracy & 83.17 & 86.34 & 91.71 & 98.05 & \underline{99.02} & \textbf{99.51}\\
         \hline
          Precision & 57.40 & 66.69 & 89.30 & \underline{92.70} & 88.50 & \textbf{99.69}\\
         \hline
          Recall & 49.34 & 67.65 & 79.41 & 93.45 & \underline{98.98} & \textbf{99.09}\\
         \hline
          F1 & 48.17 & 65.82 & 78.84 & 93.06 & \underline{99.23} & \textbf{99.36}\\
         \bottomrule
    \end{tabular}}
    \caption{\small{Classification performance comparison.}}
    \label{tab:Classification}
\end{table}

\noindent\textbf{Cell Type Classification.} 
To evaluate the discriminative power and generalization ability of the learned representations, we conducted classification experiments. 
Tab.~\ref{tab:Classification} shows that~\methodname~outperforms the baseline models on~\emph{human pancreatic cells} dataset across four metrics, including accuracy and F1-score.
This demonstrates ~\methodname's superior performance in identifying cellular heterogeneity and the robustness and effectiveness in cell type classification. 

\section{Conclusion}
\textbf{\methodname} integrates dual augmentation module, siamese fusion module, and optimal transport clustering to enhance representation learning while preserving biological relevance. 
Experimental results on seven datasets demonstrate that~\methodname~not only outperforms nine baselines on scRNA-seq clustering but also alleviates the representation collapse issue common in GNN-based approaches.  
In addition, we conduct biological analyses, including cell type annotation and classification, underscoring the~\methodname's potential as a powerful tool for advancing single-cell transcriptomics. 

\clearpage
\section*{Acknowledgements}
This work is partially supported by the National Natural Science Foundation of China (Grant No. 92470204 and 62406306), the National Key Research and Development Program of China (Grant No. 2024YFF0729201), and the Science and Technology Development Fund (FDCT), Macau SAR (file no. 0123/2023/RIA2, 001/2024/SKL).

~~\\
~~\\
~~\\
~~\\
\appendix
\large \textbf{Supplementary Materials for~\methodname}
\section{Data Details}
\subsection{Characteristic of scRNA-seq Data}
\label{supplementary_scRNA-seq}
Data generated from large-scale parallel scRNA-seq experiments forms a high-dimensional, high-sparsity gene expression matrix, where rows represent cells and columns correspond to genes. Most cell information is embedded in the non-zero expression values, with nearly 90\% of the data being zero. These datasets are further characterized by high dropout rates, substantial noise, and complex structures, complicating the process of representation learning and necessitating advanced modeling techniques.

Technical biases inherent in transcriptome sequencing amplify data noise, leading to interference in the extraction of meaningful biological signals. The sparsity of the data not only introduces computational challenges, but also blurs cellular boundaries, reducing the precision of downstream clustering and annotation tasks. In addition, the growing coverage of genes increases the complexity of intercellular relationships, making it more difficult to accurately capture subtle interactions and similarities between cells.

In this context, methods that effectively capture intercellular relationships, minimize noise interference, and enhance the interpretability of scRNA-seq data are critical for improving clustering outcomes and enabling reliable biological insights.

\subsection{Experimental Dataset}
\label{supplementary_dataset}
We evaluated the performance of the proposed~\methodname~on seven real scRNA-seq datasets by conducting extensive experiments. 
Tab.~\ref{tab:dataset} summarizes the brief information of these real datasets. 
These datasets originate from various sequencing methods. 
Of the seven datasets, three are derived from mouse samples and four from human samples. 
These datasets encompass a diverse array of cell types, such as retina, lung, liver, kidney, pancreas, and peripheral blood mononuclear cells.
\begin{table*}[!h]
    \centering
    \resizebox{\textwidth}{!}{
    \begin{tabular}{lccccc}
       \toprule
       \textbf{Datasets} & \textbf{Sequencing method} & \textbf{Sample size} & \textbf{No. of genes} & \textbf{No. of groups} & \textbf{Sparsity rate(\%)} \\
       \midrule
       Shekhar mouse retina cells~\cite{shekhar2016comprehensive} & Drop-seq & 27499 & 13166 & 19 & 93.33 \\        
       Macosko mouse retina cells~\cite{macosko2015highly} & Drop-seq & 14653 & 11422 & 39 & 88.34 \\
       QS mouse lung cells~\cite{tabula2018single} & Smart-seq2 & 1676 & 23341 & 11 & 89.08 \\
       CITE-CMBC~\cite{zheng2017massively}& 10X genomics & 8617 & 2000 & 15 & 93.26 \\ 
       Human liver cells~\cite{macparland2018single} & 10X genomics & 8444 & 4999 & 11 & 90.77 \\ 
       Human kidney cells~\cite{young2018single} & 10X genomics & 5685 & 25125 & 11 & 92.92 \\
       Human pancreas cells~\cite{muraro2016single} & CEL-seq2 & 2122 & 19046 & 9 & 73.02 \\
       \bottomrule
    \end{tabular}
    }
    \caption{Summary of the scRNA-seq datasets}
    \label{tab:dataset}
\end{table*}


\section{Abbreviations}

Tab.~\ref{tab:abbreviations} lists all abbreviations.

\begin{table}[h]
    \centering
    \begin{tabular}{ll}
        \toprule
        \textbf{Abbreviation} & \textbf{Full Form} \\
        \midrule
        scRNA-seq & single-cell RNA sequencing \\
        ZINB & zero-inflated negative binomial \\
        GNN & graph neural network \\
        SOTA & state-of-the-art \\
        KL & Kullback-Leibler \\
        ACC & accuracy \\
        NMI & normalized mutual information \\
        ARI & adjusted rand index \\
        GEA & gene expression autoencoders \\
        CGA & cell graph autoencoders \\
        CCR & cell correlation refinement \\
        LCR & latent correlation refinement \\
        \bottomrule
    \end{tabular}
    \caption{List of Abbreviations}
    \label{tab:abbreviations}
\end{table}

\section{Method Details}
\subsection{Notation Summary}
\label{supplementary_notation}
Tab.~\ref{tab:notation} summarizes the notations of all characters. 
\begin{table}[!h]
    \centering

    \resizebox{\linewidth}{!}{
    \begin{tabular}{ll}
    \toprule
    \textbf{Notations} & \textbf{Meaning} \\
    \midrule$\mathbf{X} \in \mathbb{R}^{N \times D}$ & Preprocessed matrix \\
    $\mathbf{A} \in \mathbb{R}^{N \times N}$ & Adjacency matrix (Cell graph) \\
    $\mathbf{I} \in \mathbb{R}^{N \times N}$ & Identity matrix \\
    $\mathbf{D} \in \mathbb{R}^{N \times N}$ & Degree matrix \\
    $\widetilde{\mathbf{A}} \in \mathbb{R}^{N \times N}$ & Normalized adjacency matrix \\
    $\mathbf{A}^r \in \mathbb{R}^{N \times N}$ & Edge-removed adjacency matrix \\
    $\mathbf{A}^d \in \mathbb{R}^{N \times N}$ & Graph diffusion adjacency matrix \\
    $\widehat{\mathbf{X}} \in \mathbb{R}^{N \times D}$ & Reconstructed attribute matrix \\
    $\widehat{\mathbf{A}} \in \mathbb{R}^{N \times N}$ & Reconstructed adjacency matrix \\
    $\mathbf{Z} \in \mathbb{R}^{N \times d}$ & Clustering-oriented latent embedding \\
    $\mathbf{Z}^{C} \in \mathbb{R}^{N \times d}$ & Latent embedding of GEA \\
    $\mathbf{Z}^{G} \in \mathbb{R}^{N \times d}$ & Latent embedding of CGA \\ 
    $\mathbf{R}^{\mathcal{C}} \in \mathbb{R}^{N \times N}$ & Refined cell correlation matrix \\
    $\mathbf{R}^{\mathcal{G}} \in \mathbb{R}^{d \times d}$ & Refined latent correlation matrix \\
    $\mathbf{Q} \in \mathbb{R}^{N \times C}$ & Soft assignment distribution \\
    $\mathbf{P} \in \mathbb{R}^{N \times C}$ & Target distribution \\
    \bottomrule
    \end{tabular}}
    \caption{Notation summary}
    \label{tab:notation}
\end{table}

\subsection{Data Augmentation}
\label{supplementary_data_augmentation}
Tab.~\ref{tab:sup_aug} shows the overview of data augmentations in this work. 
\begin{table}[!h]
    \centering

    \resizebox{\linewidth}{!}{
    \begin{tabular}{ccc}
        \toprule
        \textbf{Data augmentation} & \textbf{Type} & \textbf{Operation Object}\\
        \midrule
        Feature perturbation & Feature-space & Nodes\\
        Edge perturbation & structure-space & Edges\\
        Graph diffusion & structure-space & Edges\\
         \bottomrule
    \end{tabular}}
    \caption{Overview of data augmentations.}
    \label{tab:sup_aug}
\end{table}

\subsection{Gene expression and Cell Graph Autoencoders}
\label{supplementary_overall_ae}
As shown in the framework diagram of~\methodname, the encoder part adopts a Siamese network structure. 
With parameter sharing, each branch is equipped with gene expression antoencoders (GEAs) and cell graph autoencoders (CGAs) to respectively extract genes and cells information. 
In the decoder part, we use clustering-oriented latent embedding $\mathbf{Z}$ and normalized adjacency matrix $\widetilde{\mathbf{A}}$ as the input and reconstruct the inputs of the two sub-networks through the decoders of GEA and CGA, respectively. 
Unlike conventional autoencoders, GEA's decoder has three output layers, modeling highly sparse and overdispersed scRNA-seq data with a ZINB-based distribution. 
Furthermore, CGA utilizes a symmetric architecture for reconstructing both the weighted attribute matrix and the adjacency matrix. This approach enhances CGA's ability to capture both structural and attribute features of the graph, leading to more precise and comprehensive information reconstruction and a substantial enhancement in the network's generalization capability. 

\noindent\textbf{Gene Expression Autoencoders.} 
\label{supplementary_fae}
Gene expression autoencoders are proposed to learn the representative embedding of scRNA-seq expression through stacked three layers of dense networks in both the encoder and decoder. 
Taking perturbated attribute matrix  $\widetilde{\mathbf{X}}$ as inputs, we denote the feature encoder and decoder functions as:
\begin{equation}
    \begin{aligned}
        & \mathbf{Z}_{G}=f_{enc}\left(W_{GEA}\mathbf{X}+b\right) \\
        & \mathbf{H}=f_{dec}\left(W_{GEA}^{'}\mathbf{Z}_{G}+b^{'}\right),
    \end{aligned}
    \label{equ:feature_ae}
\end{equation}
where $b$ and $b^{'}$, respectively, represent the bias vectors of the encoder and the decoder, $W_{GEA}$ and $W_{GEA}^{'}$ are the weights of layers learned from network training.

\noindent\textbf{Cell Graph Autoencoder.}
\label{supplementary_sgae}
Cell graph autoencoders are designed to model complex relationships and interactions among cells, using perturbated attribute matrix  $\widetilde{\mathbf{X}}$ and the edge perturbation adjacency matrix $\mathbf{A}^r$ (or graph diffusion adjacency matrix $\mathbf{A}^d$) as input, which are essential for a deeper understanding of cellular diversity.
This network requires to reconstruct both the weighted attribute matrix and the adjacency matrix simultaneously. Moreover, a layer in the encoder and decoder is formulated as: 
\begin{equation}
\begin{aligned}
&\mathbf{Z}_{C}^{(l)}=\sigma\left(\mathbf{A}^r \mathbf{Z}_{C}^{(l-1)} W_{CGA}^{(l)}\right), \\
&\widehat{\mathbf{Z}}_{C}^{(h)}=\sigma\left(\mathbf{A}^r \widehat{\mathbf{Z}}_{C}^{(h-1)} \widehat{W}_{CGA}^{(h)}\right),
\end{aligned}
\end{equation}
where $W_{CGA}^{(l)}$ and $\widehat{W}_{CGA}^{(h)}$, respectively, denote the learnable parameters of $l$-th encoder layer and $h$-th decoder layer. $\sigma$ is a non-linear activation function, such as ReLU or Tanh.  

To minimize both the reconstruction loss functions over GEA and CGA, our model is designed to minimize a hybrid loss function~\cite{tu2021deep}: 
\begin{equation}
\scalebox{0.75}{$
\begin{aligned}
\mathcal{L}_{REC} & = \mathcal{L}_{GEA}+\mathcal{L}_{CGA}\\
& = \frac{1}{2N}\|\widetilde{\mathbf{X}}-\widehat{\mathbf{X}}\|_F ^2+\left(\frac{1}{2N}\|\widetilde{\mathbf{A}}\widetilde{\mathbf{X}}-\mathbf{Z}\|_F ^2+\eta\left(\frac{1}{2N}\|\widetilde{\mathbf{A}}-\widehat{\mathbf{A}}\|_F ^2\right)\right), 
\label{equ:loss_rec}
\end{aligned}
$}
\end{equation}
where, $\eta$ is a pre-defined hyper-parameter. 
By minimizing Eq.~\ref{equ:loss_rec},  is termed to minimize the reconstruction loss over node attributes and graph structure adopted. 

\noindent\textbf{Denoising ZINB model-based Feature Decoder.}
\label{supplementary_zinb}
The ZINB distribution models highly sparse and overdispersed count data~\cite{eraslan2019single}. 
To explain the characteristics of scRNA-seq data, we assume that the data follow ZINB distribution, namely:
\begin{equation}
\scalebox{0.8}{$
\begin{aligned}
\mathrm{NB}(\mathbf{X}_{ij}; \mu_{ij}, \theta_{ij})=\frac{\Gamma(\mathbf{X}_{ij}+\theta_{ij})}{\Gamma(\theta_{ij})}\left(\frac{\theta_{ij}}
{\theta_{ij}+\mu_{ij}}\right)^{\theta_{ij}}
\left(\frac{\mu_{ij}}{\theta_{ij}+\mu_{ij}}\right)^{\mathbf{X}_{ij}},
\label{equ:nb}
\end{aligned}
$}
\end{equation}

\begin{equation}
\small
\operatorname{ZINB}(\mathbf{X}_{ij} ; \pi_{ij}, \mu_{ij}, \theta_{ij})=\pi_{ij} \delta_0(\mathbf{X}_{ij})+(1-\pi_{ij}) \mathrm{NB}(\mathbf{X}_{ij} ; \mu, \theta),
\label{equ:zinb}
\end{equation}
where dropout ($\pi_{ij}$) are the probability of dropout events, mean ($\mu_{ij}$) and dispersion ($\theta_{ij}$) parameters, respectively, represent the negative binomial component. 

The decoder has three output layers to estimate parameters $\pi$, $\mu$ and $\theta$, which are
\begin{equation}
\begin{aligned}
&\hat{\pi}=\operatorname{sigmoid}\left(W_\pi \cdot \mathbf{H}\right), \\
&\hat{\mu}=diag\left(s_i\right)\times\exp \left(W_\mu \cdot \mathbf{H}\right), \\
&\hat{\theta}=\exp \left(W_\theta \cdot \mathbf{H}\right).
\end{aligned}
\label{equ:zinb_parameters}
\end{equation}
Similarly, $W_\pi$, $W_\mu$ and $W_\theta$ are the corresponding weights, the size factors $s_i$ is the ratio of the total cell count to the median $S$. 
Finally, we minimize the following overall negative likelihood loss of the  ZINB model-based autoencoder, namely
\begin{equation}
\mathcal{L}_{Z I N B }=-log\left(Z I N B\left(\mathbf{X}_{ij} \mid \hat{\pi}_{ij}, \hat{\mu}_{ij}, \hat{\theta}_{ij}\right)\right)
\label{loss_zinb}
\end{equation}

\subsection{Similarity Measurements}
\noindent{\textbf{Embedded Points and Clustering Center.}}
\label{sec:appendix_similarity_measure_1}
We use the Student’s t-distribution~\cite{van2008visualizing} to measure the similarity between embedded points and the clustering center.
The Student’s t-distribution~\cite{van2008visualizing} is a popularly used kernel to measure the similarity between the embedded point $h_{i}$ and the clustering center $c_{j}$. It is defined as follows,
\begin{equation}
    q_{ij} = \frac{(1 + \|h_{i} - c_{j}\|^{2}/\varepsilon)^{-\frac{1 + \varepsilon}{2}}}{\sum_{j^{'}}(1 + \|h_{i} - c_{j^{'}}\|^{2}/\varepsilon)^{-\frac{1 + \varepsilon}{2}}},
\label{equ:kl_qij}
\end{equation}
where $h_{i}=f(\bm{x}_{i}) \in \mathbf{H}$ corresponds to $\bm{x}_{i} \in \widetilde{\mathbf{X}}$ after embedding, $\varepsilon$ is the degrees of freedom of the Student’s t-distribution, and $q_{ij}$ can be interpreted as the probability of assigning sample $i$ to cluster $j$ (i.e., a soft assignment). 
We define $Q=\left[q_{i j}\right]$ as the distribution of the assignments of all samples.

\noindent{\textbf{Data Representation and Clustering Center.}}
\label{sec:appendix_similarity_measure_2}
We use the target distribution distance ~\cite{bo2020structural} to measure the similarity between data representation and the clustering center. Specifically, we calculated a target distribution $P=\left[p_{i j}\right]$:
\begin{equation}
    p_{ij}=\frac{q_{ij} ^2 / f_j}{\sum{q_{ij^{'}}^2 / f_{j^{'}}}},
\end{equation}
where $f_j=\sum_{i} a_{ij}$ are soft cluster frequencies. In the target distribution $P$, each assignment in $Q$ is squared and normalized to enhance assignment confidence~\cite{bo2020structural}. 

\section{Experimental Details}
\subsection{Evaluation Metrics}
\label{supplementary_evaluation_metrics}
Given the knowledge of the ground truth class assignments $U$ and our clustering algorithm assignment $V$ on $n$ data points.

\noindent \textbf{Accuracy (ACC):} The ACC is defined to measure the best matching between two cluster assignments $U$ and $V$. If there is a data point $i$, $l_i$ is the ground truth label, and $u_i$ is the clustering algorithm's assignment, then the ACC is defined as:
\begin{equation}
    \text{ACC}=\max_{m}\frac{\sum_{i=1}^n 1 {l_i=m(u_i)}}{n},
\label{equ:ca}
\end{equation}
here, $m$ ranges over all possible one-to-one mapping between $U$ and $V$, the best mapping with respect to $m$ can be efficiently found using the Hungarian algorithm~\cite{kuhn1955hungarian}.

\noindent \textbf{Normalized Mutual Information (NMI):} The NMI~\cite{strehl2002cluster} is a function that measures the consistency between the predicted and true labels of $n$ cells. Specifically,
\begin{equation}
\small
    \text{NMI}=\frac{\sum_{i=1}^{C_U} \sum_{j=1}^{C_V} \frac{\mid U_i\bigcap V_j\mid}{n}\text{log}\frac{n\mid U_i\bigcap V_j\mid}{\mid U_i\mid\mid V_j\mid}}{\text{mean}(-\sum_{i=1}^{C_U} \mid U_i\mid \text{log} \frac{\mid U_i\mid}{n},-\sum_{j=1}^{C_V} \mid V_j\mid \text{log} \frac{\mid V_j\mid}{n})}.
\label{equ:nmi}
\end{equation}

\noindent \textbf{Adjusted Rand Index (ARI):} The ARI~\cite{vinh2009information} is a function that evaluates the similarity of the two assignments ignoring permutations. Given a set $X$ of size $n$ and two clustering results $U = \{u_1, u_2,\cdots, u_r\}$, $V = \{v_1, v_2,\cdots, v_s\}$, we denote $n_{i j}=\left|U_i \cap V_j\right|, i=1,2, \ldots, r, j = 1, 2,\cdots, s$. We also let $a_i=\sum_{i=1}^r n_{i j}$, $b_j=\sum_{j=1}^s n_{i j}$. Then we evaluate:
\begin{equation}
\text{ARI} = \frac{\sum_{i,j} \binom{n_{i,j}}{2} - \left[ \sum_i \binom{a_i}{2} \sum_j \binom{b_j}{2} \right] \binom{n}{2}}{\left[ \sum_i \binom{a_i}{2} \sum_j \binom{b_j}{2} \right]/ 2 - \left[ \sum_i \binom{a_i}{2} \sum_j \binom{b_j}{2} \right]/\binom{n}{2}}.
\label{equ:ari}
\end{equation}

\noindent \textbf{Macro Precision:} 
It calculates the precision for each class (i.e., the proportion of correctly predicted positive instances for each class out of all instances predicted as that class), and then averages the precision over all classes.
\begin{equation}
\text{Macro Precision} = \frac{1}{N} \sum^{N}_{i=1} \frac{TP_{i}}{TP_{i} + FP_{i}}
\label{Macro_Precision}
\end{equation}
where \(TP_{i}\) is the count of true positives for class \(i\), \(FP_{i}\) is the count of false positives for class \(i\), and \(N\) represents the total number of classes. 
This metric is particularly useful for assessing the performance of classification models, especially in multi-class scenarios.

\noindent \textbf{Macro Recall:} 
It calculates the recall for each class (i.e., the proportion of correctly predicted positive instances for each class out of all actual positive instances), and then averages the recall over all classes.
\begin{equation}
\text{Macro Recall} = \frac{1}{N} \sum^{N}_{i=1} \frac{TP_{i}}{TP_{i} + FN_{i}}
\label{Macro_Recall}
\end{equation}
where \(TP_{i}\) is the number of true positives for the \(i^{th}\) class, and \(FN_{i}\) is the number of false negatives for the \(i^{th}\) class.
It indicates the model's average ability to retrieve positive instances for each class.

\noindent \textbf{Macro F1:} 
It calculates the F1 score for each class (i.e., the harmonic mean of precision and recall for each class), and then averages the F1 scores over all classes.
\begin{equation}
\text{Macro F1} = \frac{1}{N} \sum^{N}_{i=1} \frac{2 * Precision_{i} * Recall_{i}}{Precision_{i} + Recall_{i}}
\label{Macro_F1}
\end{equation}
It is a comprehensive metric used to evaluate the overall performance of a model, particularly in the presence of imbalanced classes.

\subsection{Baseline Methods}
\label{supplementary_baseline_details}
Ten clustering competing methods are used in our experimental design. 
Among those approaches, pcaReduce~\cite{vzurauskiene2016pcareduce} and SUSSC~\cite{wang2021suscc} stand out as early traditional clustering techniques. 
Additionally, we consider four deep neural network-based clustering methods, namely DEC~\cite{xie2016unsupervised}, scDeepCluster~\cite{tian2019clustering}, scDCC~\cite{tian2021model}, and scNAME~\cite{wan2022scname}, which train an autoencoder to obtain representations followed by applying a clustering algorithm.
scDSC~\cite{gan2022deep}, scGNN~\cite{wang2021scgnn}, and scCDCG~\cite{xu2024sccdcg} are representative GNN-based clustering methods that incorporate both node attributes and structural information to obtain latent representation for clustering.
\begin{itemize}
    \item \textbf{Early Traditional Clustering Methods:}
    \begin{itemize}
        \item \textbf{pcaReduce}~\cite{vzurauskiene2016pcareduce} merges PCA with K-means clustering and iteratively amalgamates clusters based on relevant probability density functions.
        \item \textbf{SUSSC}~\cite{wang2021suscc} employs stochastic gradient descent together with a nearest neighbor search strategy to achieve efficient and accurate clustering.
    \end{itemize}
    
    \item \textbf{Deep Neural Network-based Clustering Methods:}
    \begin{itemize}
        \item \textbf{DEC}~\cite{xie2016unsupervised} is deep-embedded clustering, employing neural networks to learn both feature representations and cluster assignments.
        \item \textbf{scDeepCluster}~\cite{tian2019clustering}, a single-cell model-based deep embedded clustering method, which simultaneously learns feature representation and clustering via explicit modeling of scRNA-seq data generation.
        \item \textbf{scDCC}~\cite{tian2021model} encodes prior domain knowledge as constraint information, integrating it effectively into the representation learning process using a novel loss function designed to enhance clustering accuracy.
        \item \textbf{scNAME}~\cite{wan2022scname} integrates a mask estimation task alongside a neighborhood contrastive learning framework, optimizing both denoising and clustering procedures for scRNA-seq data.
    \end{itemize}

    \item \textbf{Deep Structural Clustering Methods:}
    \begin{itemize}
        \item \textbf{scDSC}~\cite{gan2022deep} learns data representations using the ZINB model-based autoencoder and GNN modules, and uses mutual supervised strategy to unify these two architectures.
        \item \textbf{scGNN}~\cite{wang2021scgnn} utilizes GNNs and three multi-modal autoencoders to formulate and aggregate cell-cell relationships and models heterogeneous gene expression patterns using a left-truncated mixture Gaussian model.
        \item \textbf{scCDCG}~\cite{xu2024sccdcg} is a clustering framework for scRNA-seq data that combines cut-informed graph embedding, self-supervised learning with optimal transport, and autoencoder-driven feature learning to fully leverage complex intercellular relationships.
    \end{itemize}
\end{itemize}

\subsection{Classification Model}
\label{supplementary_classification}
The classification model adopts a four-layer linear structure with ReLU activation functions and batch normalization between each layer. Specifically, we employed a Multilayer Perceptron (MLP) architecture consisting of input, two hidden, and output layers 
, ultimately predicting cell types. The dataset was meticulously partitioned into training and testing sets with an 8:2 ratio to ensure robust generalization during model training. Optimization objectives were guided by the CrossEntropyLoss function, and optimization was performed using the Adam optimizer with an initial learning rate set to 1e-2. Furthermore, we employed the ReduceLROnPlateau scheduler to dynamically adjust the learning rate based on loss minimization, utilizing a decay factor of 0.9 to expedite convergence to the optimal solution. Throughout the training phase, we meticulously conducted 100 epochs with validation checks performed every 10 epochs to meticulously monitor training progress and performance.

\section{Implementation Details}
\label{supplementary_implementation_details}
\noindent\textbf{Training Procedure.}
The training process of~\methodname~includes three steps: (i) we pre-train the GEA and CGA independently for at least 50 epochs by minimizing the reconstruction loss functions; (ii) both subnetworks are integrated into a united framework for another 100 epochs; (iii) Under the guidance of Eq.14 in the main text, we train the entire network for at least 50 epochs until convergence.
As the number of cell clusters $k$ is unknown in real applications, we adopt the K-means method over the consensus clustering embedding $\mathbf{Z}$ to determine the optimal value of $k$. 
The code is available at the link: \url{https://github.com/XPgogogo/scSiameseClu}. 

\noindent\textbf{Parameters Setting.} 
In practice, we implement scSiameseClu in Python 3.7 based on PyTorch. 
For competing methods, we reproduce their source code by following the setting of the original literature and present the average results.
For our proposed method, we adopt DCRN~\cite{liu2022deep} as our backbone network. 
For all datasets, the trade-off parameter $\gamma$ was set to 1e3. The hyper-parameters $\beta$, $\lambda$, $\eta$, and $\varepsilon$ were initially set to 0.5, 0, 5, 0.1, and 1, respectively.
Specific values were assigned to $\alpha$ for different datasets: 0.32 for \emph{CITE-CMBC}, 0.25 for \emph{Human liver cells}, 0.5 for \emph{Human kidney cells}, and 0.1 for all other datasets. Similarly, $\rho$ was set to 100 for \emph{CITE-CMBC}, 50 for \emph{Human kidney cells}, and 10 for the remaining datasets. The value of $\sigma$ was adjusted to 5 for \emph{Human kidney cells}, 1 for \emph{QS mouse lung cells}, and 2 for the rest.
The Adam optimizer was employed for optimization, with learning rates tailored to each dataset: 5e-3 for \emph{Human kidney cells}, 5e-4 for \emph{QS mouse lung cells}, and 1e-4 for all other datasets.
To ensure the accuracy of the experimental results, we conducted each experiment 10 times with random seeds on all the datasets and reported the mean and variance of the results. 

\section{Experiment Results}
\label{supplementary_results}
\subsection{Additional Quantitative Analysis}
\label{supplementary_overall_perference_analysis}
By further analyzing the experimental results in Tab.1 in the main text, we can draw the conclusion:
\methodname~significantly improves upon existing techniques, demonstrating superior representation learning capabilities on scRNA-seq data while successfully addressing the issue of representation collapse. Specifically, by thoroughly extracting intercellular high-order structural information and integrating gene expression attributes,~\methodname~effectively learned discriminative features within the scRNA-seq data. 
Our approach improves data interpretability and robustly supports the discovery of complex biological relationships between cells, thereby facilitating further advancements in biological research.

\begin{figure*}[!t]
\centering
\subfloat[DEC]{
\includegraphics[width=0.5\textwidth]{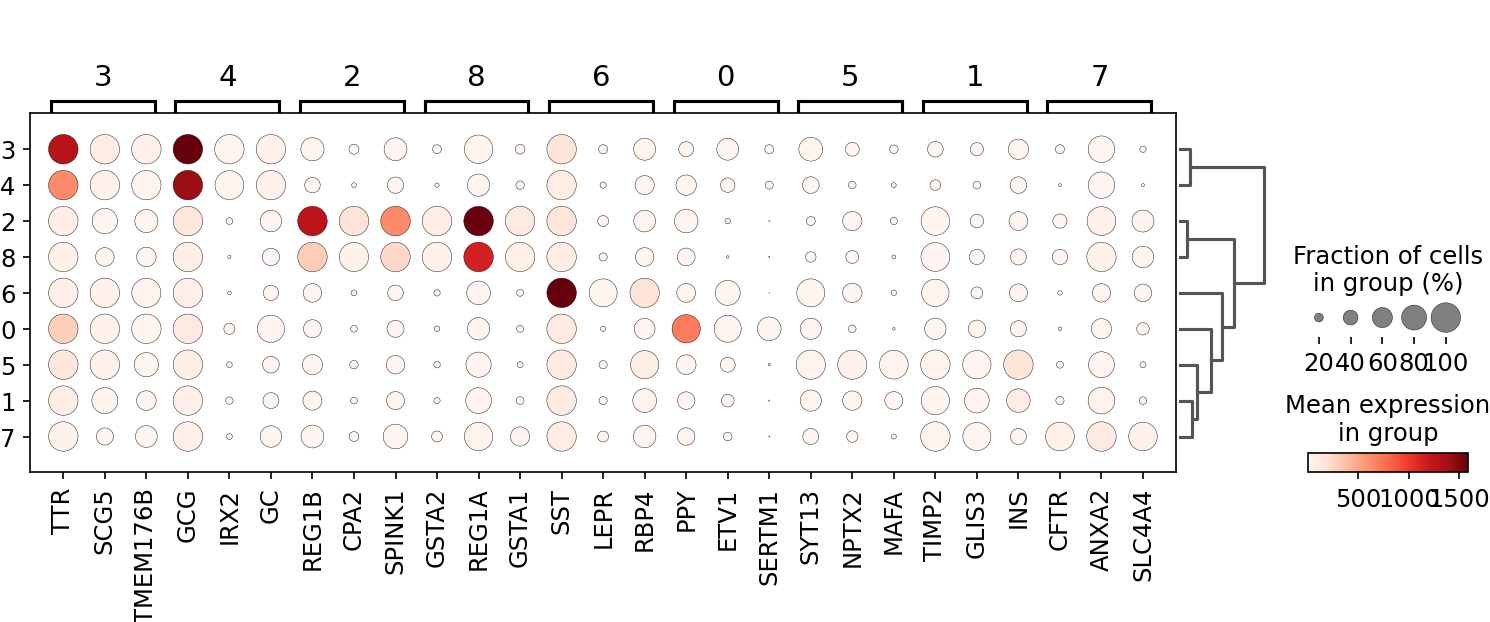}
}
\subfloat[scDeepCluster]{
\includegraphics[width=0.5\textwidth]{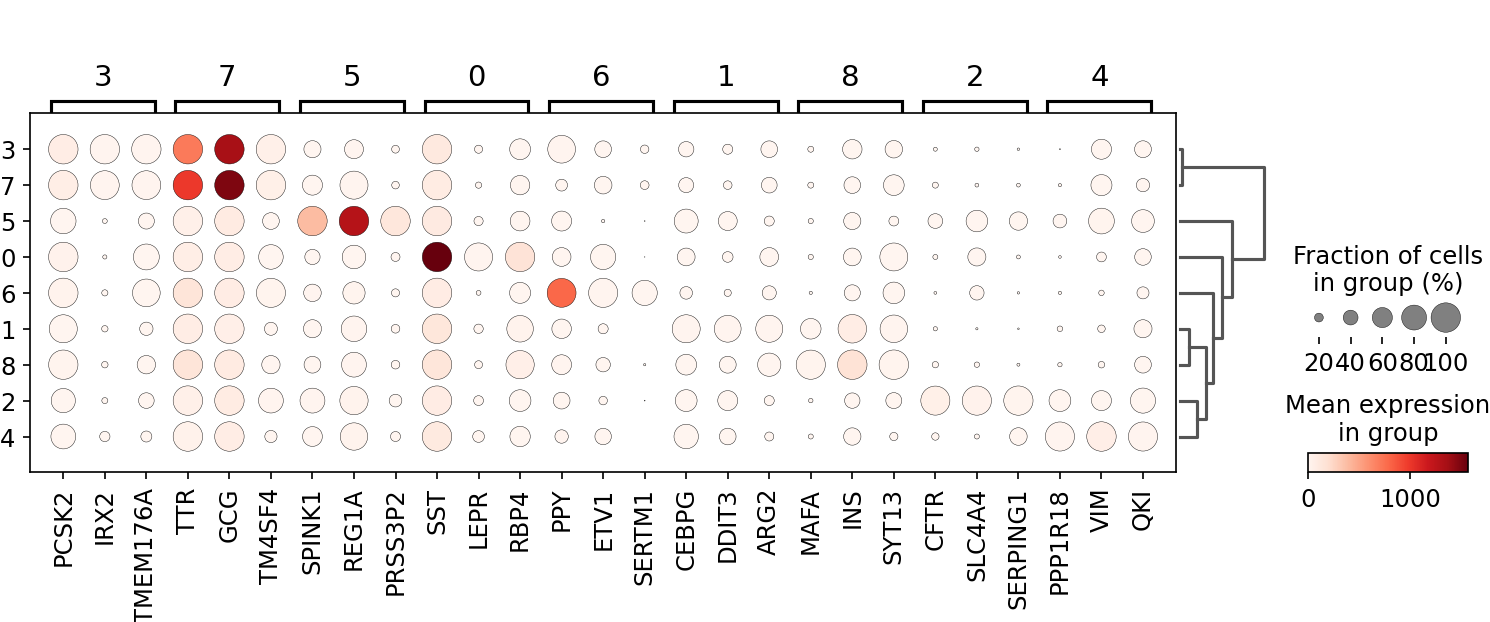}
}
\\
\subfloat[scDCC]{
\includegraphics[width=0.5\textwidth]{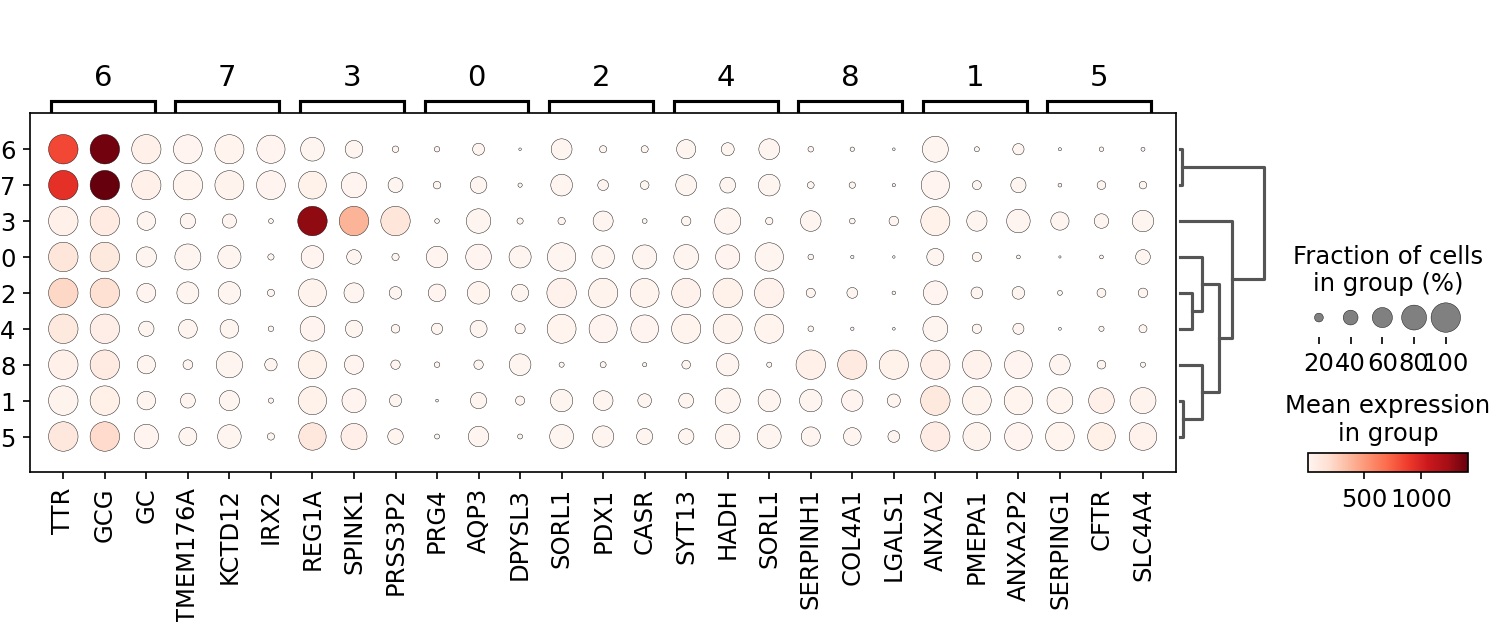}
}
\subfloat[scCDCG]{
\includegraphics[width=0.5\textwidth]{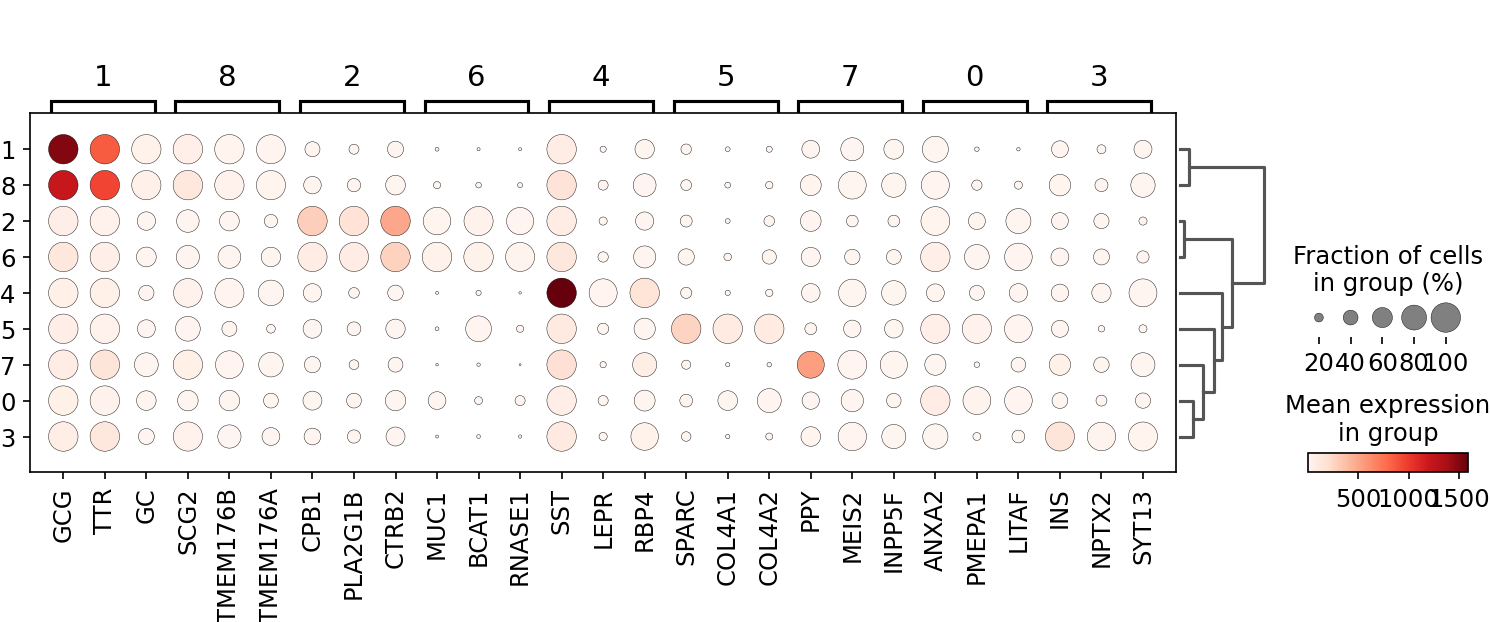}
}

\caption{Dot plot of the top three DEGs for each cell type identified by four baseline models using the~\emph{human pancreatic cells} dataset.}
\label{fig:DEG_meruo_supply}
\end{figure*}

\subsection{Additional Biological Analysis.}
\label{supplementary_bioanalysis}

\begin{figure*}[!t]
\centering
\subfloat[scDeepCluster]{
\includegraphics[width=0.18\textwidth]{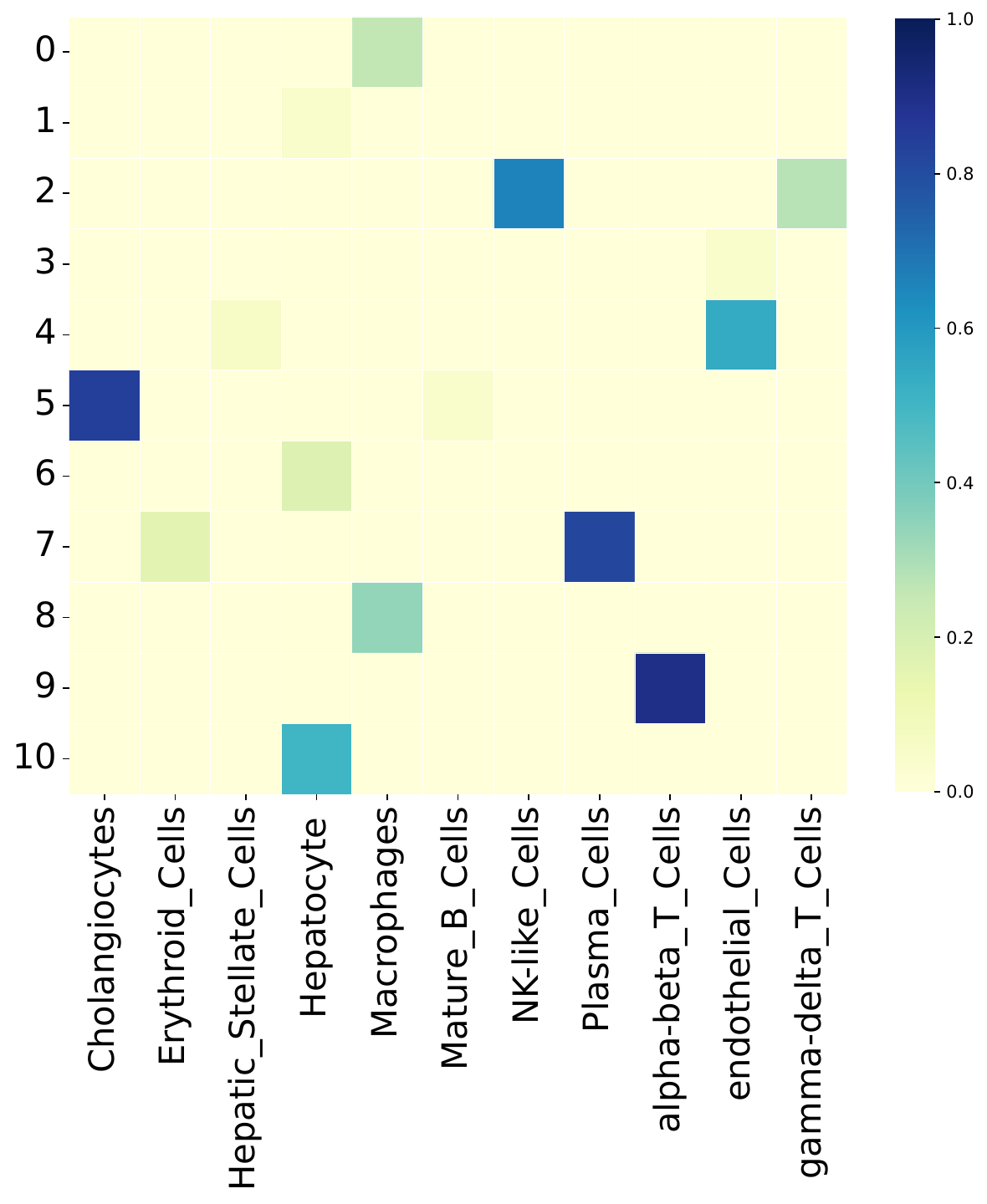}
}
\subfloat[scDCC]{
\includegraphics[width=0.18\textwidth]{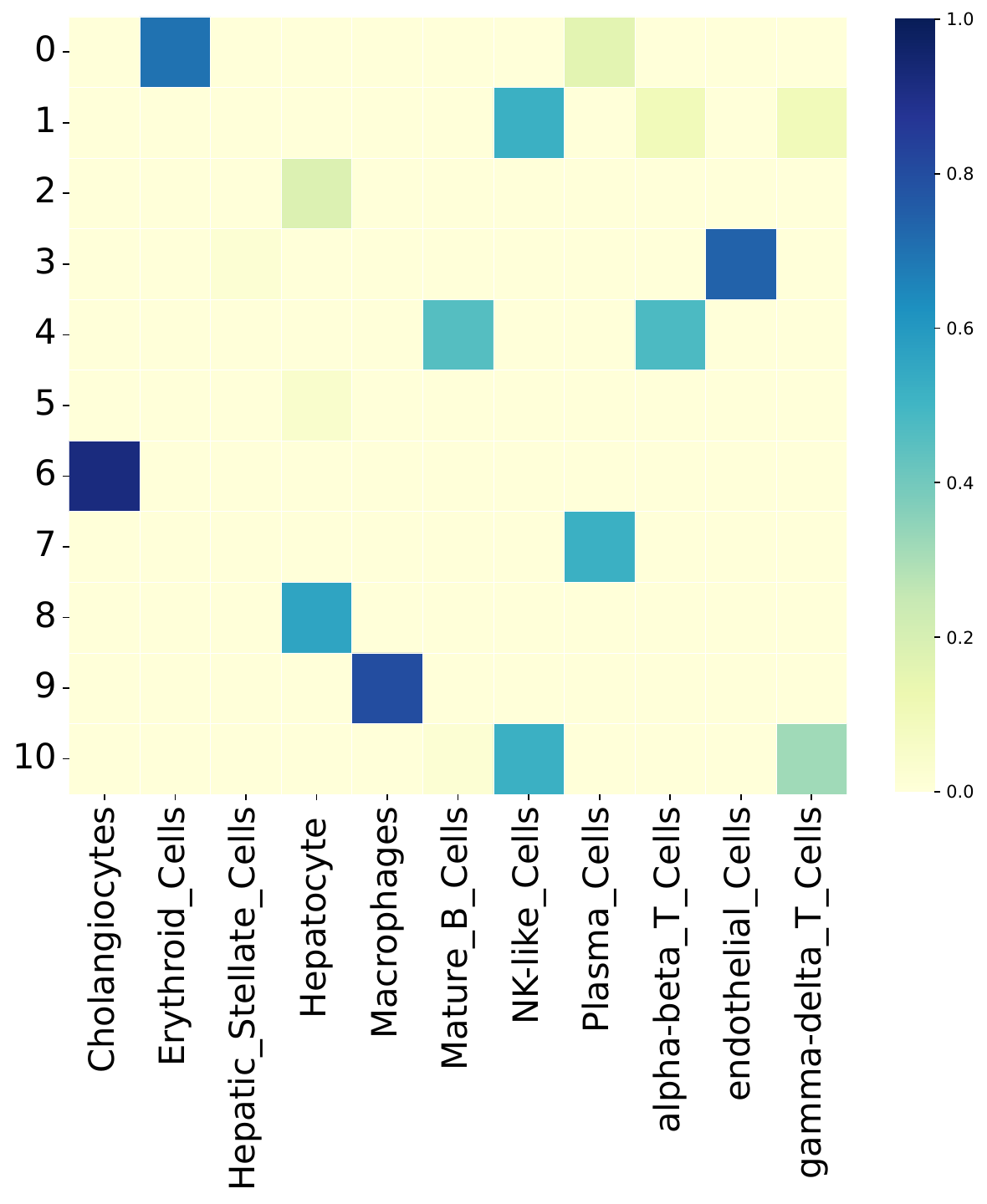}
}
\subfloat[scDSC]{
\includegraphics[width=0.18\textwidth]{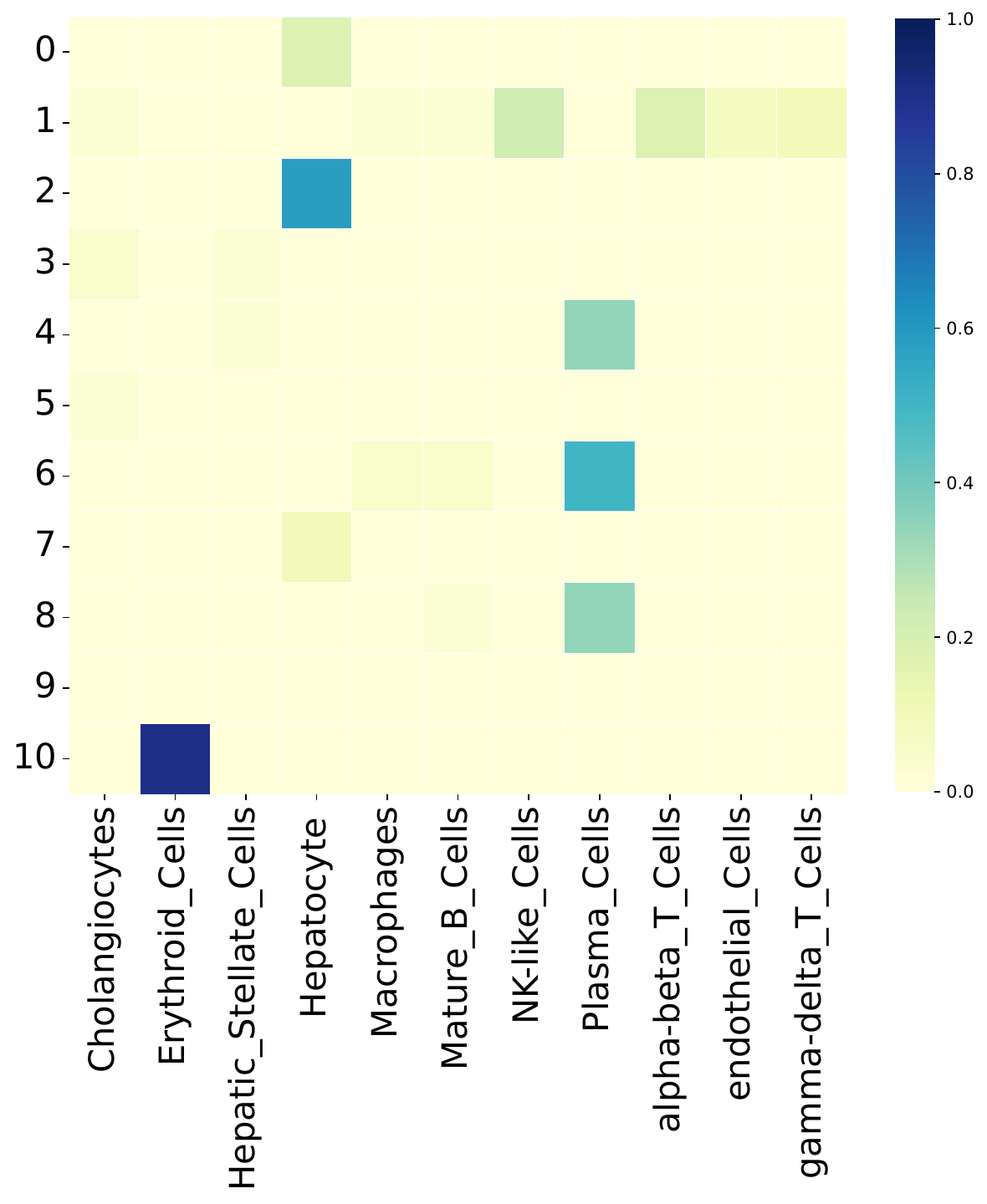}
}
\subfloat[scCDCG]{
\includegraphics[width=0.18\textwidth]{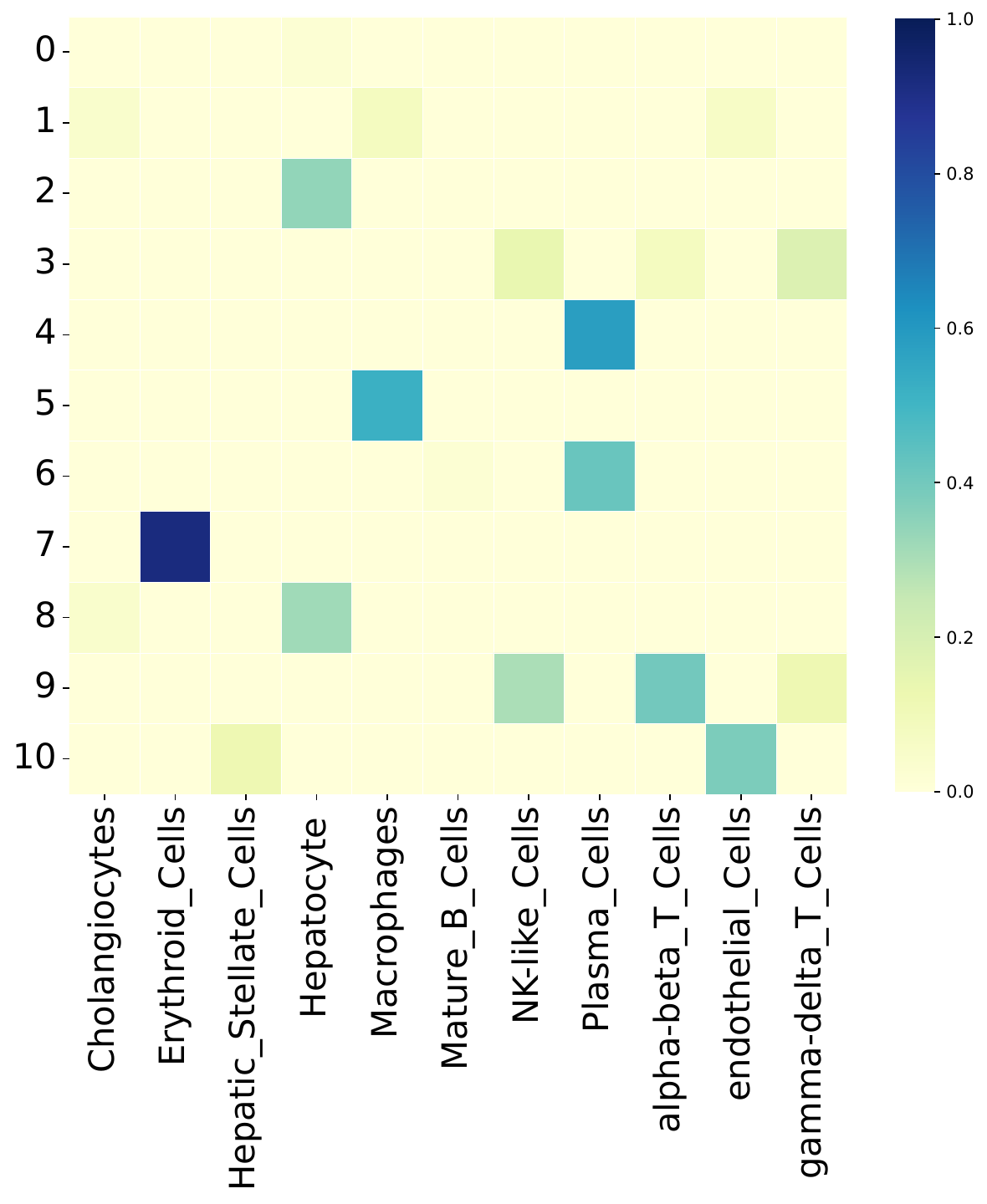}
}
\subfloat[\methodname]{
\includegraphics[width=0.18\textwidth]{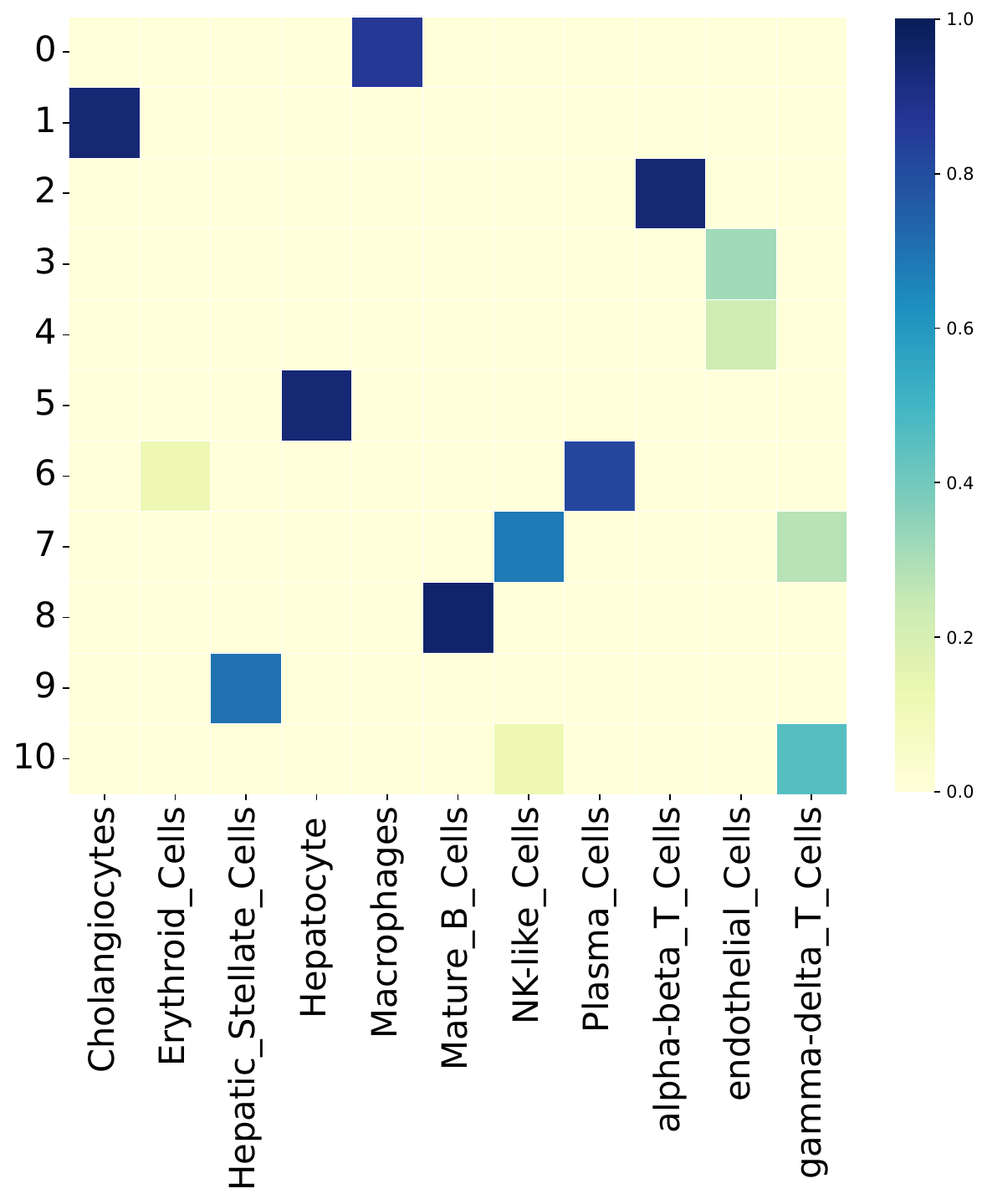}
}
\caption{Cell type annotation: overlap of the top 50 DEGs in clusters from five methods versus gold standard cell types useing the~\emph{human liver cells} dataset. (similarity = overlapping DEGs/50).}
\label{fig:CTA_sonya}
\end{figure*}

\begin{figure}[!t]
    \centering
    \includegraphics[width=1\linewidth]{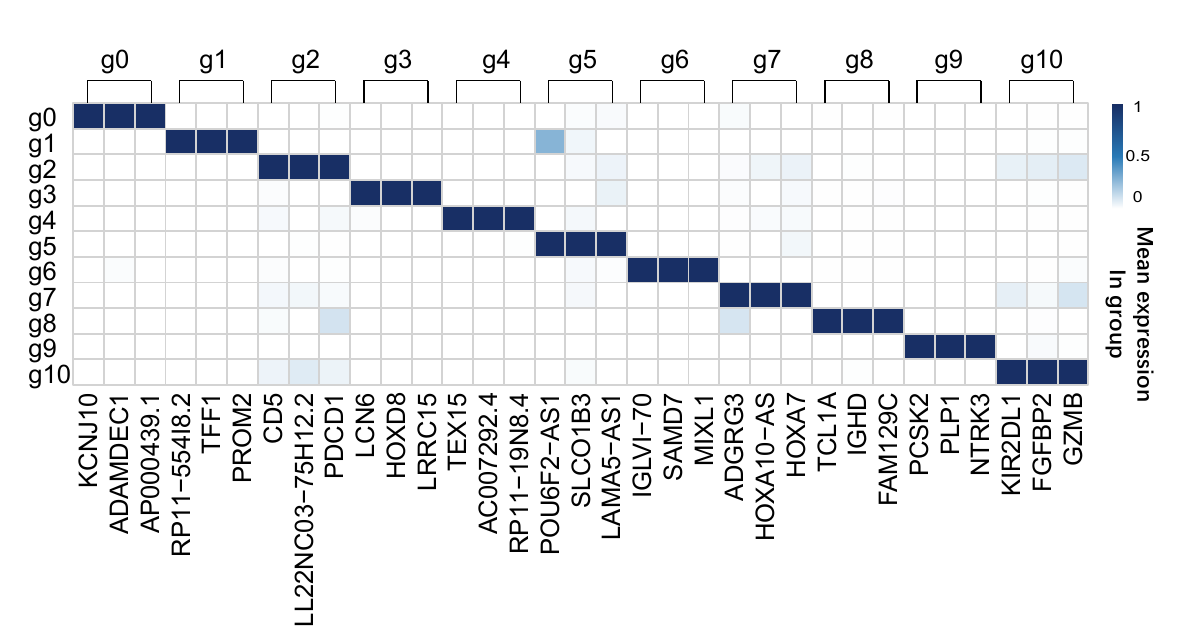}
    \caption{Matrix plot of the top 3 DEGs for each cell type by ~\methodname useing the~\emph{human liver cells} dataset.}
    \label{fig:DEG_sonya}
\end{figure}

Based on the previous experimental results, we have validated the capability of~\methodname~to learn discriminative feature representations and achieve excellent clustering performance. 
Analyzing gene expression patterns through clustering of scRNA-seq data has proven advantageous in expediting the learning and application of various biological downstream tasks. 
Therefore, extracting meaningful biological insights from clustering results is essential for practical applications.
Here, taking~\emph{human pancreatic cells} and~\emph{Human liver cells} as examples, we further conduct biological analysis to find out whether our method is beneficial for downstream analysis, including marker gene identification, cell type annotation, and cell type classification, based on the labels predicted by the~\methodname. 

The gene expression matrix and predicted labels were utilized to identify DEGs and further recognize marker genes for each cluster.
Standard cell type annotation methods were applied to \emph{Human pancreatic cells} and \emph{Human liver cells} datasets, assigning cell types to each cluster based on established markers and cluster-specific gene expression profiles. 
We employed the "FindAllMarkers" function in the Seurat~\cite{butler2018integrating} package with default parameters to identify DEGs for each cluster, thereby obtaining DEGs within the gold standard clusters annotated with respective cell types. 
Differentially expressed genes (DEGs) are important indicators for the cell type annotations of the clusters obtained through different methods. 
To determine whether~\methodname~can accurately annotate clusters, we processed~\methodname~and each competing method identically, identifying the top 50 marker genes for each cluster and comparing their overlap with the gold standard, referred to as the similarity of DEGs between different methods and the gold standard (as shown in Fig.5 in the main text,~\methodname~could assign a cell type to a specific cluster, with most clusters showing a differential gene similarity greater than 90\%. For example, cluster 1 was identified as endothelial cells, and cluster 2 as type B pancreatic cells, etc.
scDeepCluster could not identify the cell type for cluster 1; scNAME could not identify cell types for clusters 2 and 7, nor could it assign a unique cell type to cluster 0; scGNN and scCDCG could not assign endothelial cell to a specific cluster. 

Detailed results showed that other methods, such as scDeepCluster, scNAME, scGNN, and scCDCG, failed to consistently identify specific cell types for various clusters, highlighting the superiority of our approach in accurately matching clusters with known cell types.
To further DEGs identified by the~\methodname~clustering, Fig.6 in the main text depicts a dot plot of the top 3 DEGs for each of the 9 clusters. These genes represent the most characteristic DEGs within each cluster identified by our method and can be considered as the marker genes for the respective clusters.

In further analysis of the~\emph{human pancreatic cells} dataset, we plotted the top three differentially expressed genes (DEGs) for each cell type from four baseline models (DEC, scDeepCluster, scDCC, scSDCG), as shown in Fig.~\ref{fig:DEG_meruo_supply}. 
Comparing these results with Fig.6 in the main text, we found that~\methodname~excels in identifying cell type-specific DEGs, particularly demonstrating greater consistency and reliability in the selection of highly expressed genes. This provides clear support for subsequent biological research.
Fig.~\ref{fig:CTA_sonya} and Fig.~\ref{fig:DEG_sonya} in the supplementary materials are supplemented with experimental results conducted on~\emph{Human liver cells} dataset.

Therefore, our clustering results can help researchers identify marker genes and annotate cells, which is able to further facilitate the understanding of different cell types, uncover biological differences between cells, and provide crucial insights for downstream functional analysis and mechanistic studies.
\subsection{Ablation Study}
\begin{figure}[!t]
\centering
\subfloat[Major components evaluaion]{
\includegraphics[width=0.22\textwidth]{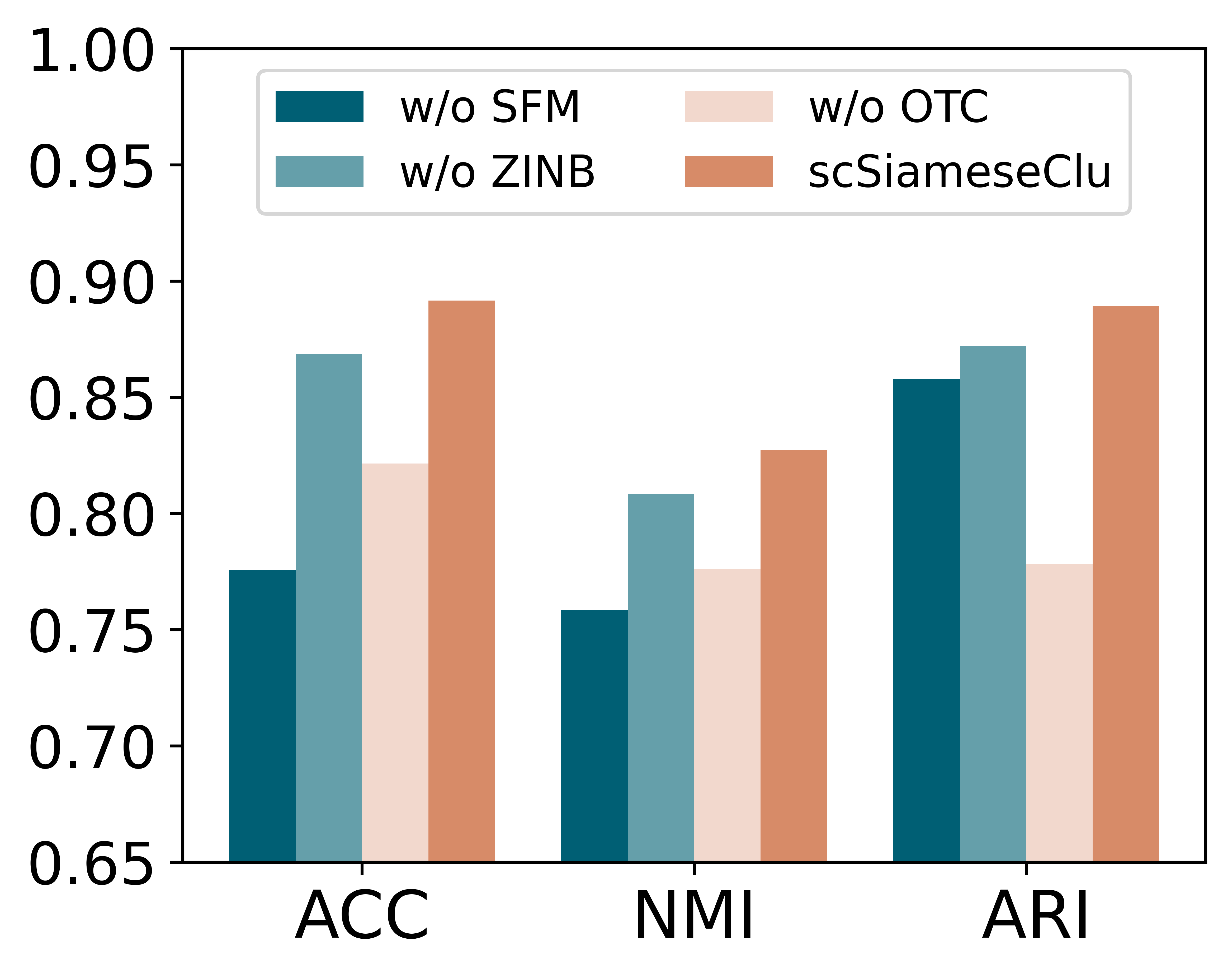}
}
\subfloat[Component Effects in SFM]{
\includegraphics[width=0.22\textwidth]{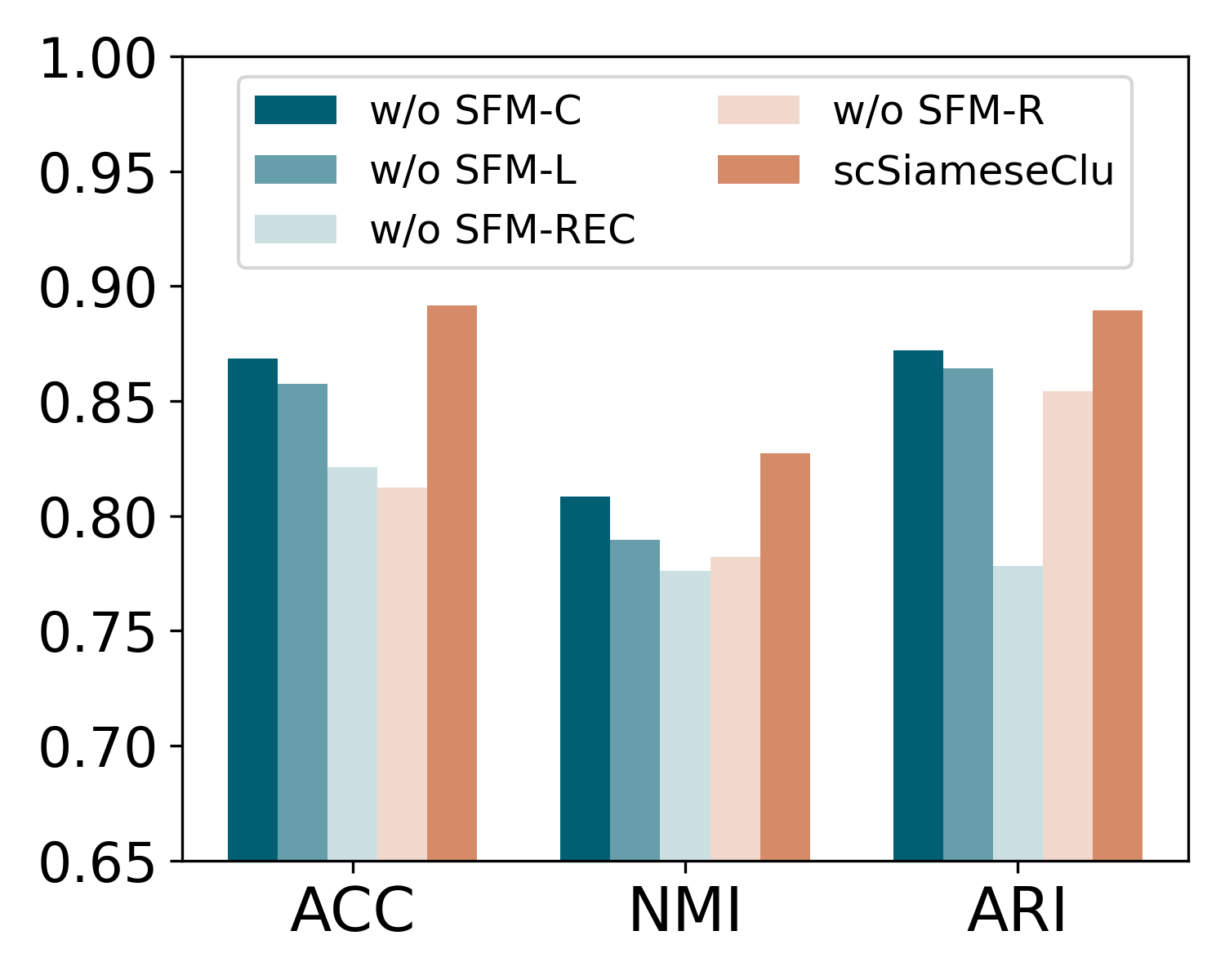}
}
\caption{Ablation study on ~\emph{Shekhar mouse retina cells} dataset.}
\label{fig:ablation}
\end{figure}

\noindent\textbf{Framework Component Study.} 
To evaluate the effectiveness of each component in ~\methodname, we perform an ablation study on  ~\emph{Shekhar mouse retina cells} dataset.  
Specifically, we denote the model without SFM loss as \emph{\methodname-w/o SFM}, while  
\emph{\methodname-w/o ZINB} and \emph{\methodname-w/o OTC} denote the model without ZINB loss and OTC loss, respectively. 
Likewise,~\methodname~refers to the complete model with all components included. 
According to Fig.~\ref{fig:ablation}(a), each component significantly enhances the model's performance, demonstrating their importance in improving efficiency and outperforming the baseline. 

\noindent\textbf{SFM Module Evaluation.} 
As shown in Fig.\ref{fig:ablation}(a), SFM module plays a significant role in improving overall performance. 
To further validate the importance of its components, we compare our method (\methodname) with its baseline through ablation experiments (Fig.~\ref{fig:ablation}(b)).  
Specifically, \emph{w/o SFM-C} and \emph{w/o SFM-L} denote the model without the cell correlation and latent correlation, respectively. 
Likewise, we denote the model without propagated regularization as \emph{w/o SFM-R}, without MSE reconstruction loss as \emph{w/o SFM-REC}.
\methodname~also all these components. 
Fig.~\ref{fig:ablation}(b) shows that each SFM component improves performance, indicating that that~\methodname~effectively integrates both gene and cell information. 

\subsection{Computational Efficiency}
The running time of scSiameseClu and seven baselines (DEC, scDeepCluster, scDCC, scNAME, scDSC, scGNN, and scCDCG). 
The results show that scSiameseClu consistently demonstrates superior computational efficiency across various datasets, while baseline methods experience varying computational burdens. The improved efficiency of scSiameseClu reduces training time and computational resource requirements, making it a viable and powerful tool for practical applications in biological data analysis. 

\begin{figure}[!t]
    \centering
    \includegraphics[width=1\linewidth]{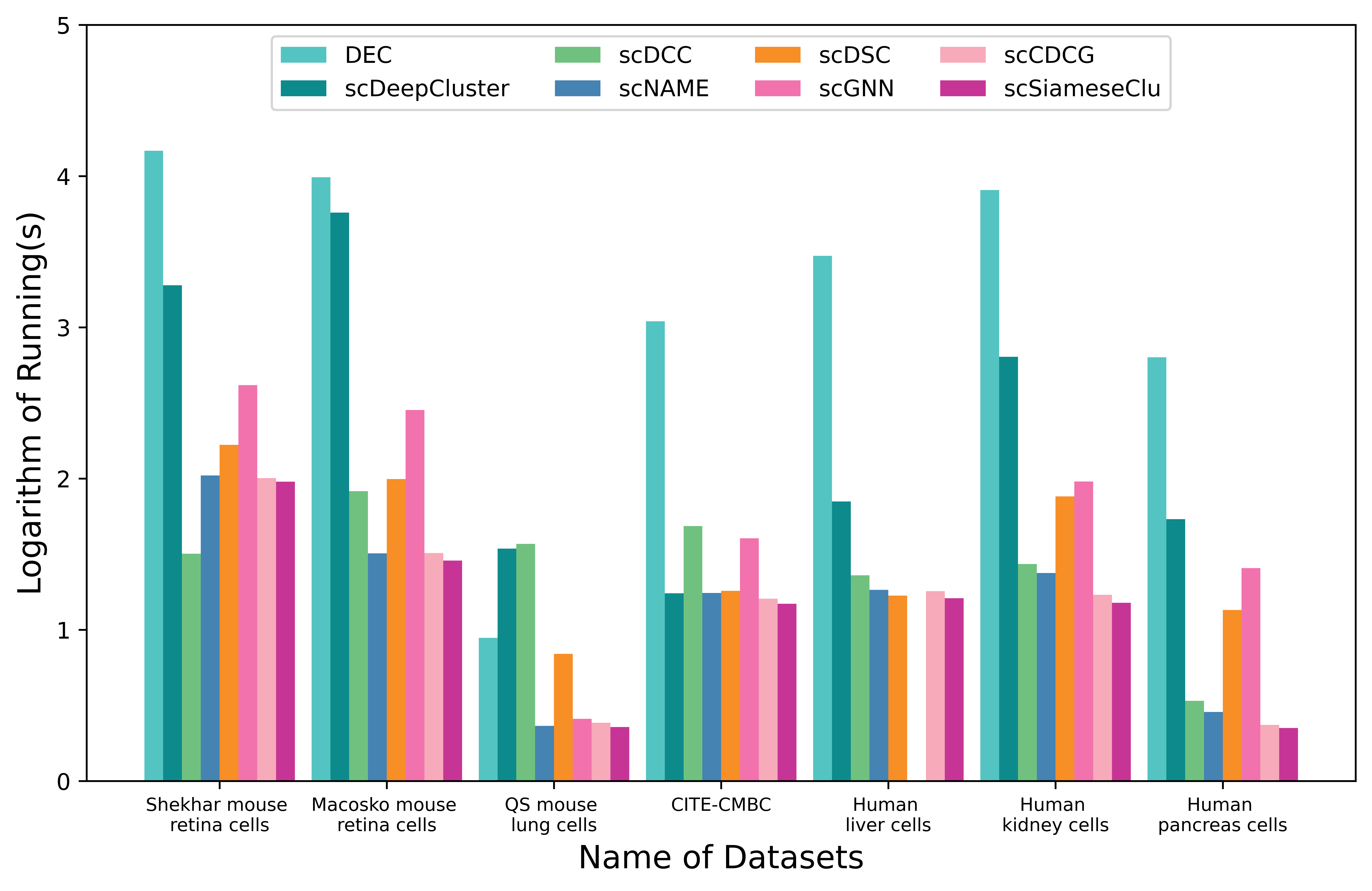}
    \caption{The running time of~\methodname~and seven baselines.}
    \label{fig:running_time_all_model}
\end{figure}

\subsection{Scalability of~\methodname}
\begin{figure}[!t]
\centering
\includegraphics[width=0.5\textwidth]{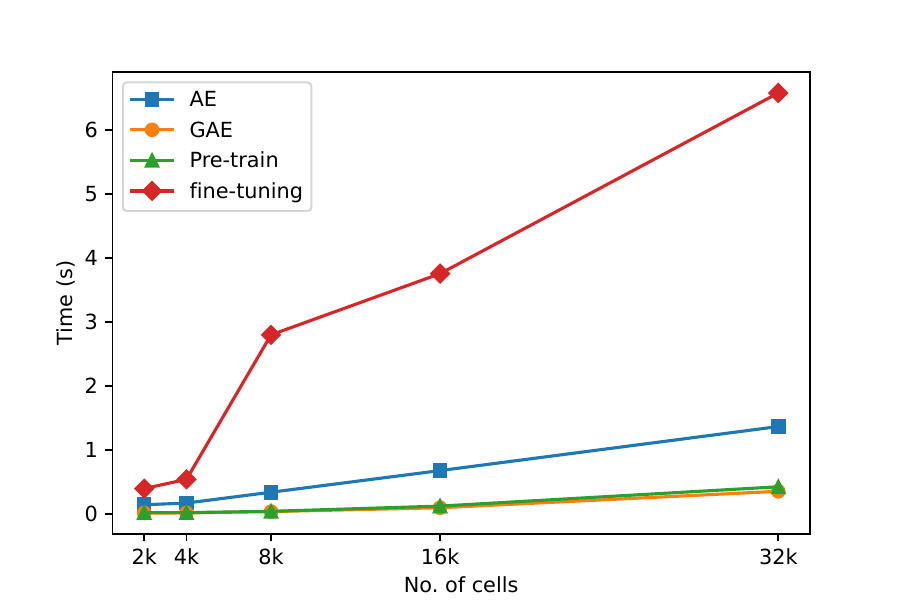}
    \caption{Applying scSiameseClu on various simulated data. }
    \label{fig:simulated_time}
\end{figure}

With the increasing number of cells being profiled in scRNA-seq experiments, there is an urgent need to develop analytical methods that can effectively handle large datasets. 
To assess the scalability of~\methodname, we summarized the runtime of ~\methodname~on simulation datasets of varying sizes. 
We designed the following simulations under extensive settings approximating different biological scenarios. 
Specifically, we applied the R package Splatter~\cite{zappia2017splatter} to simulate seven simulated datasets, repsectively, containing 2k, 4k, 8k, 16k, 32k cells with 3k genes.
Fig.~\ref{fig:simulated_time} reports the runtime for one epoch of~\methodname~at different stages (i.e., AE, GAE, pre-train, and fine-tuning) on these datasets.
It can be observed that as the number of cells increases, the runtime increases almost linearly, with the most significant increase occurring during the fine-tuning stage.


\begin{figure*}[!t]
\centering
\subfloat[$\alpha$]{
\includegraphics[width=0.24\textwidth]{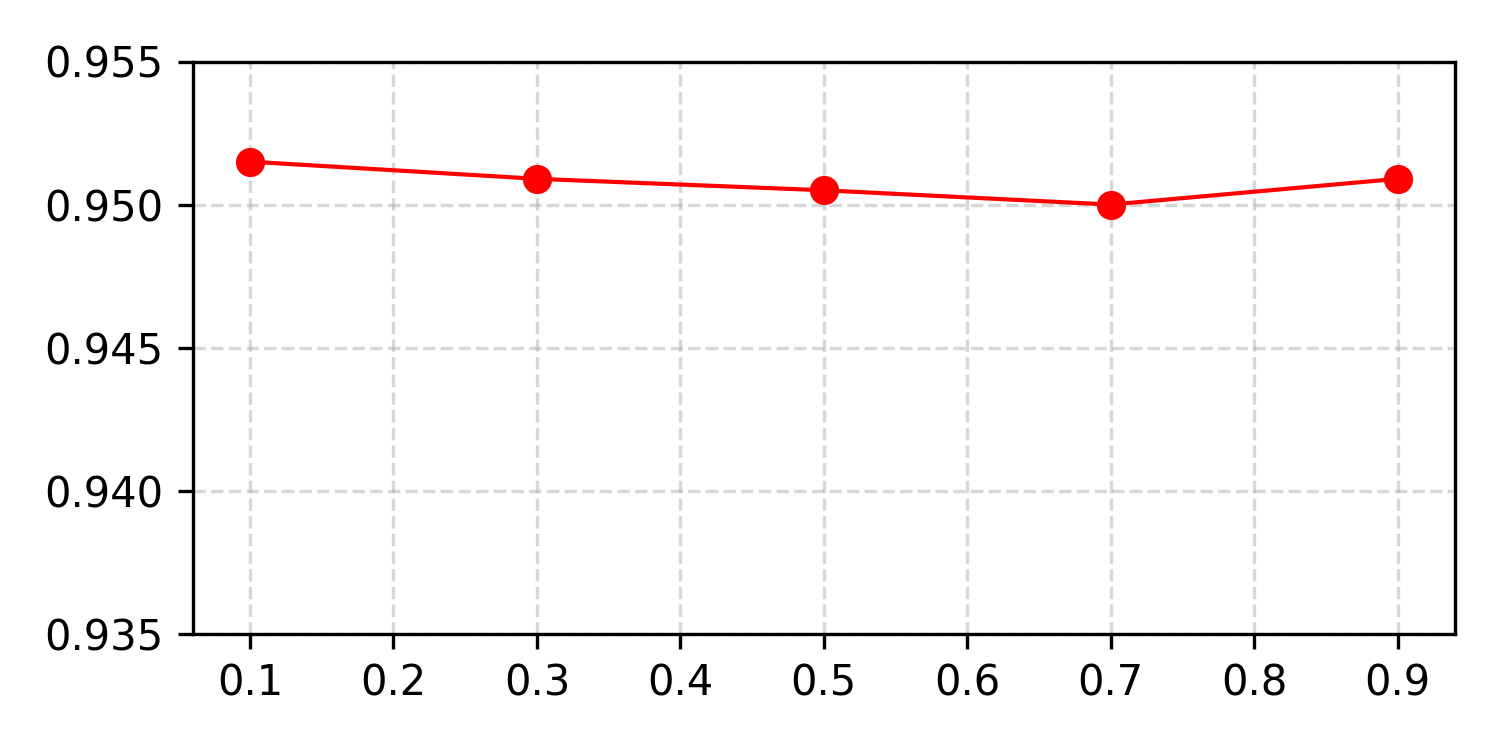}
}
\subfloat[$\beta$]{
\includegraphics[width=0.24\textwidth]{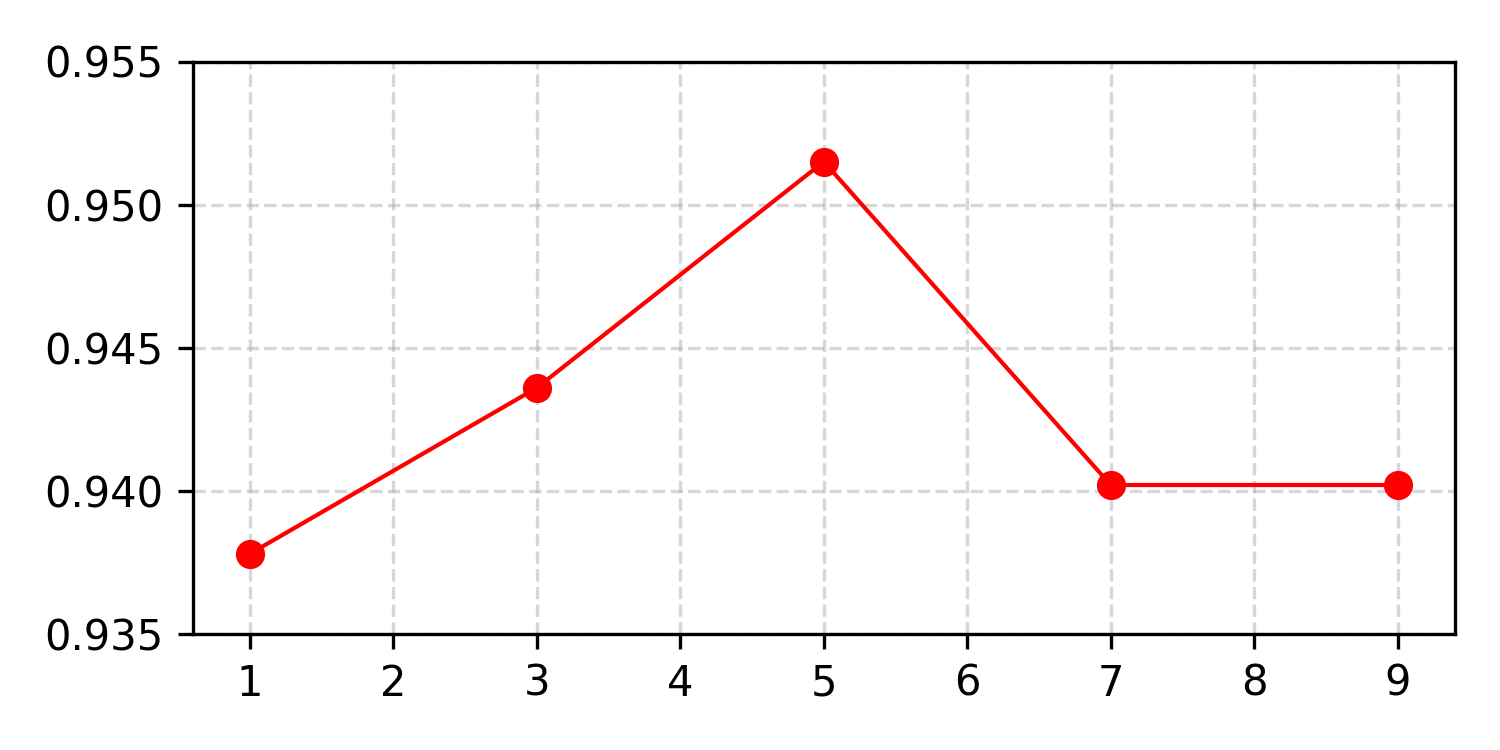}
}
\subfloat[$\gamma$]{
\includegraphics[width=0.24\textwidth]{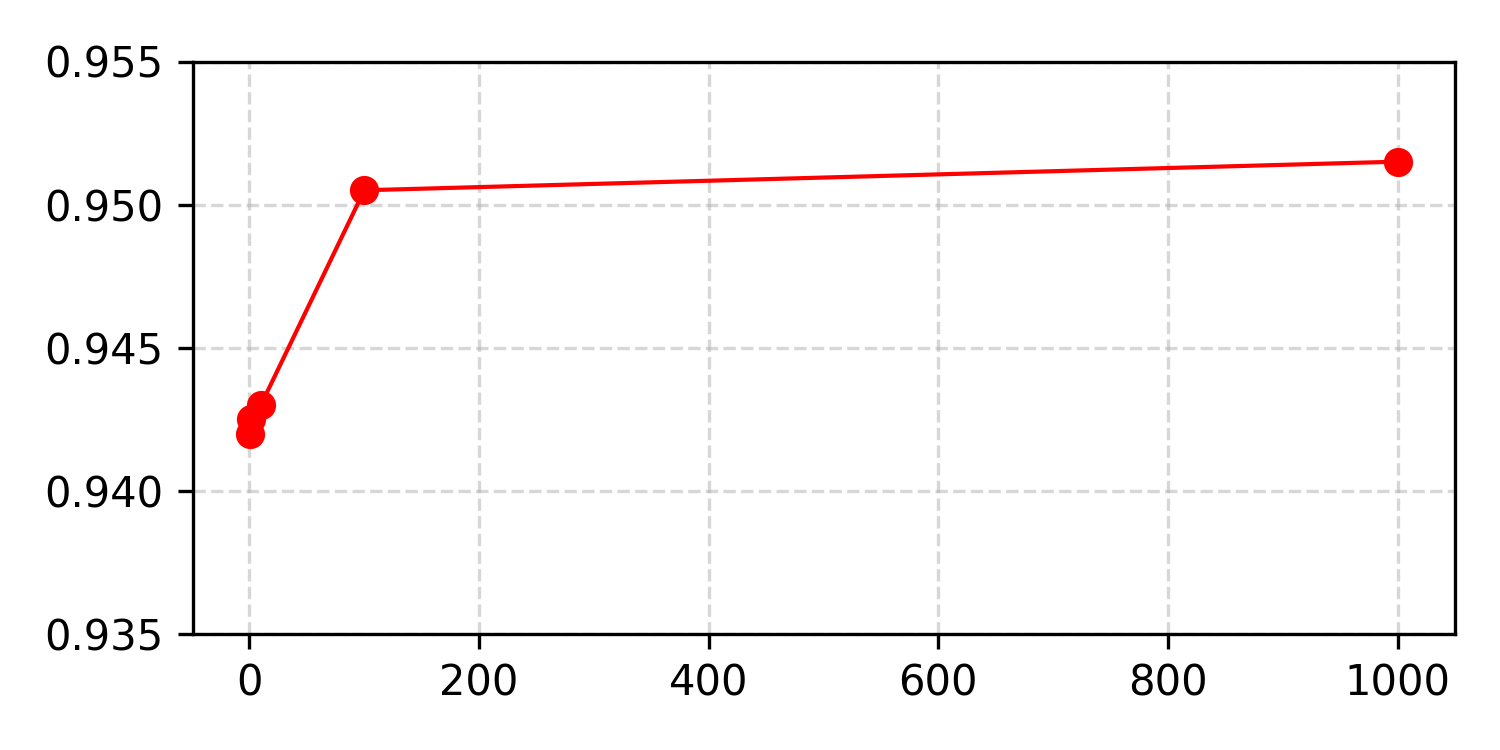}
}
\\
\vspace{-2mm}
\subfloat[$\lambda$]{
\includegraphics[width=0.24\textwidth]{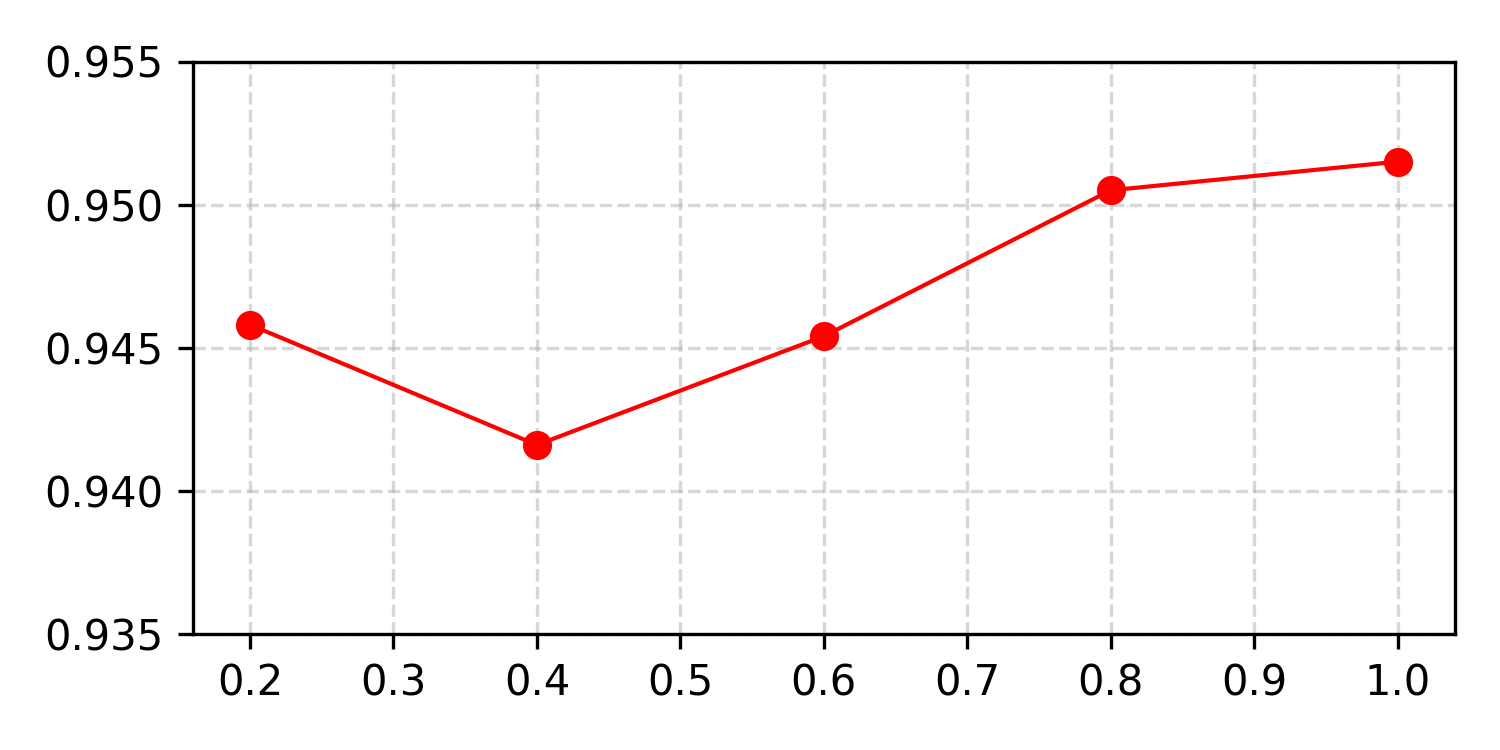}
}
\subfloat[$\eta$]{
\includegraphics[width=0.24\textwidth]{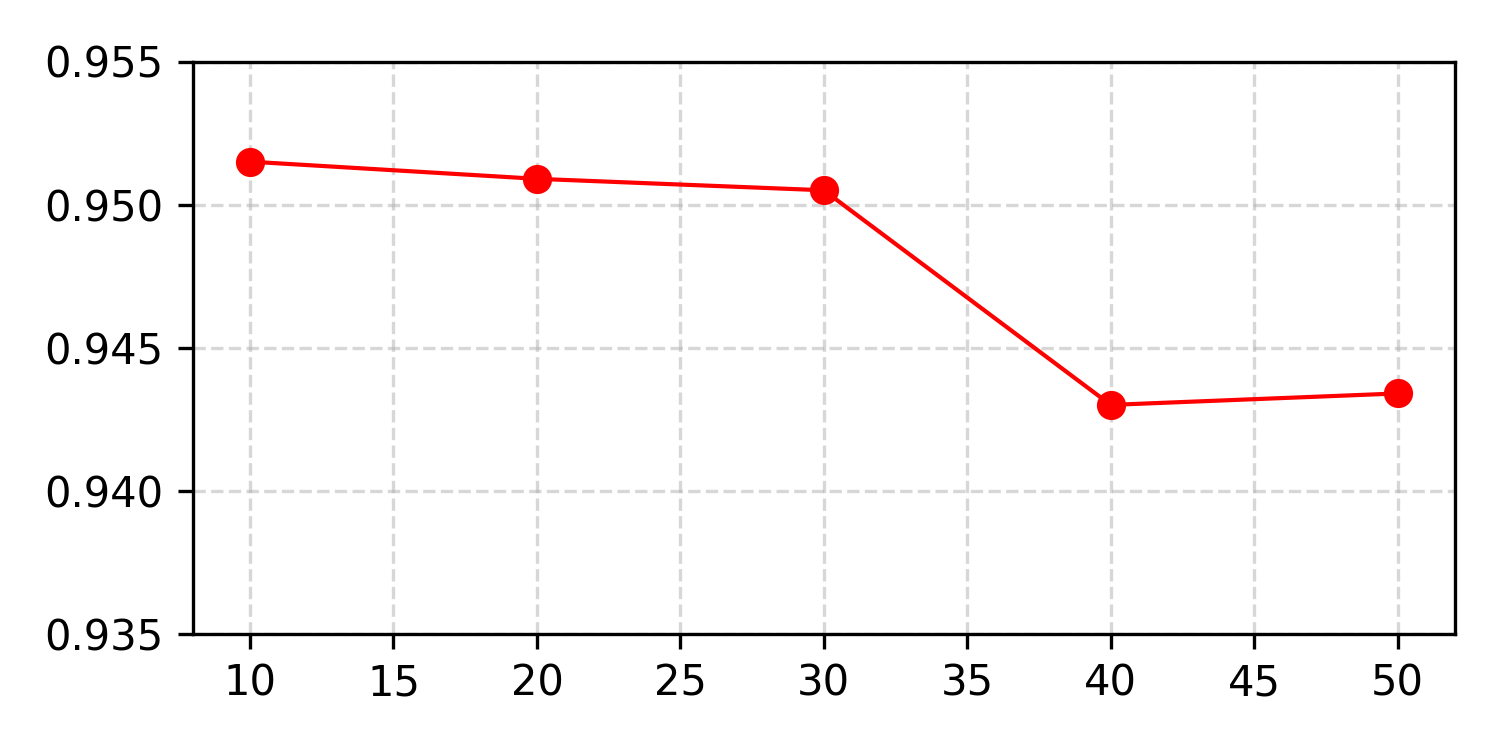}
}
\subfloat[$\sigma$]{
\includegraphics[width=0.24\textwidth]{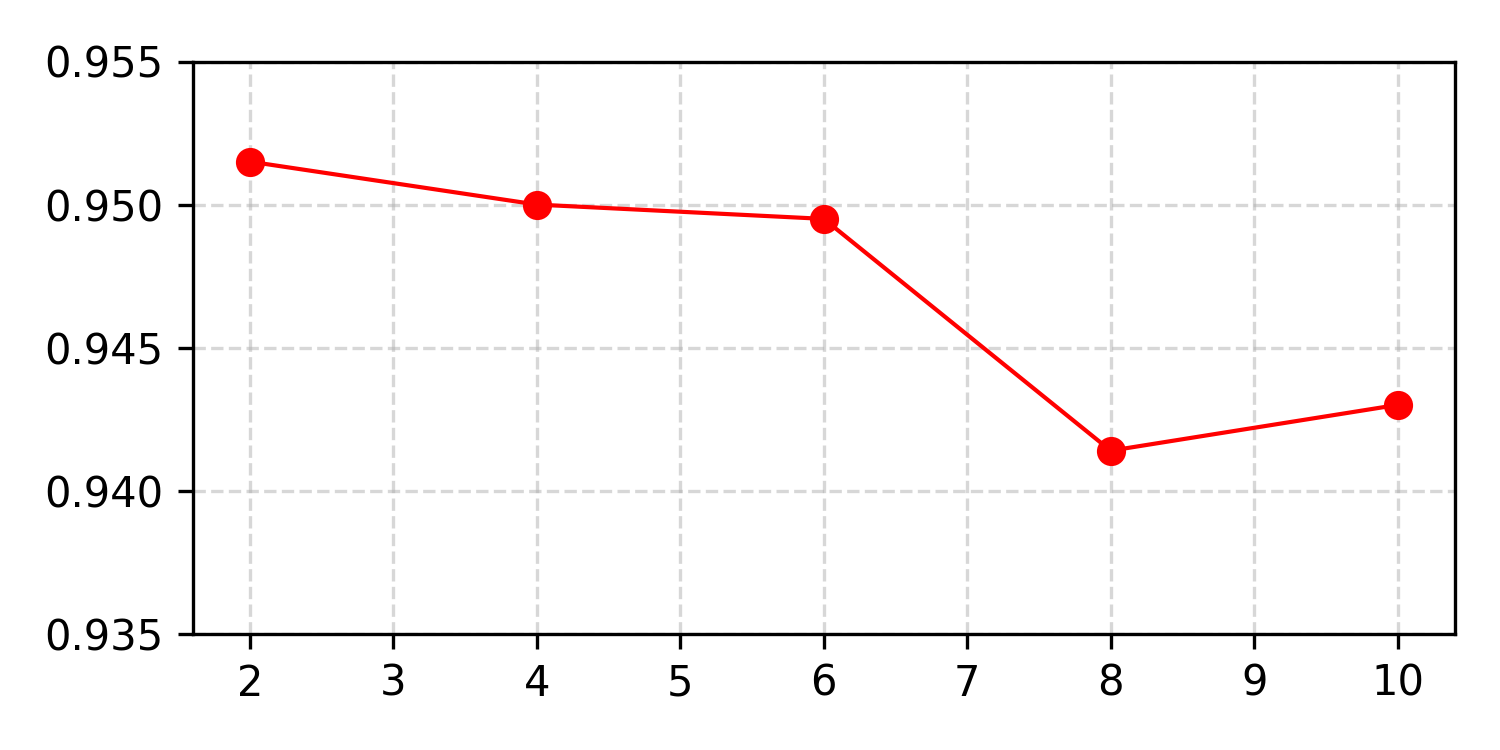}
}
\caption{Performance of~\methodname~on \emph{Human Pancreas cell} benchmark datasets w.r.t. different hyper-parameters, including $\alpha$, $\beta$, $\gamma$, $\lambda$, $\eta$, and $\sigma$.}
\label{fig: Parameter_sensitivity}
\end{figure*}
\subsection{Parameter sensitivity}
Our work introduces seven hyperparameters: $\alpha$, $\beta$, $\gamma$, $\lambda$, $\mu$, $\delta$, and $\sigma$, which balance the contributions of different modules. Here are the details:
\begin{itemize}
    \item $\alpha$: Probability of message passing through higher-order information for generating the graph diffusion matrix.
    \item $\beta$: Proportion parameter that controls the fusion ratio of $\mathbf{Z}^{E}$ and $\mathbf{Z}^{R}$; its weight is learned during training.
    \item $\lambda$: Smoothing parameter controlling cluster balance in the optimal transport strategy.
    \item $\eta$, $\delta$, and $\sigma$: Parameters controlling the contributions of different modules in the loss function. 
\end{itemize}

As shown in the Fig.~\ref{fig: Parameter_sensitivity}, we examined the impact of these hyperparameters on model performance through a sensitivity analysis using \emph{Human Pancreas cell} dataset. 
These results indicate that scSiameseClu maintains stable performance across a wide range of hyperparameter values. 
For instance, values of  between $0.1$ and $0.9$ all yielded ACC values around $0.95$, demonstrating robustness to this hyperparameter. Similarly, the other hyperparameters also show stable performance, which is critical for practical applications where tuning hyperparameters can be challenging.
To enhance the reproducibility of experiments and the generalizability of the model, we recommend initially setting a group of hyperparameters: 
$\alpha=0.1$, $\beta=1$, $\gamma=1e3$, $\lambda=5$, $\eta=10$, and $\sigma=2$. 
These values serve as starting points for hyperparameter tuning, and adjusting them based on the specific characteristics of the data is necessary when applying~\methodname~to new datasets. Although our sensitivity analysis indicates that these settings generally yield near-optimal results across various datasets, fine-tuning the parameters to accommodate different datasets helps ensure the model's stability and optimal performance.

\section{Related Work}
\label{supplementary_relatedwork}
\subsection{Traditional Clustering Algorithm}
In the realm of scRNA-seq data analysis, early traditional clustering algorithms typically employ hard clustering methods, which involve dimensionality reduction followed by clustering using algorithms like k-means or hierarchical clustering. 
Within these traditional approaches, Phenograph~\cite{levine2015data}, MAGIC~\cite{van2018recovering}, and Seurat~\cite{butler2018integrating} utilize k-nearest neighbor (KNN) graphs to simulate relationships between cells, with MAGIC being the first method to explicitly impute the entire genome in single-cell gene expression profiles. In contrast, CIDR~\cite{lin2017cidr} adopts an implicit imputation method to mitigate the impact of dropout events. 
SIMLR~\cite{wang2018simlr} and MPSSC~\cite{park2018spectral}, on the other hand, are multi-kernel spectral clustering methods that leverage multiple kernel functions to learn robust similarity measures corresponding to different information representations in the data. MPSSC is well suited to represent the sparsity of scRNA-seq data. 
Additionally, RaceID~\cite{lin2017cidr} introduces outlier detection to enhance the ability of K-means to identify subpopulation cells, while scImpute~\cite{li2018accurate} is a statistical method that aims to accurately and robustly impute dropout values in scRNA-seq data. 
However, these methods often rely on hard clustering algorithms, making it difficult to handle high-dimension data and capture complex nonlinear data features, thus making them susceptible to inherent noise in the data. 

\subsection{Deep Feature Clustering Algorithm}
Deep learning has also gained prominence in scRNA-seq clustering.
Methods like scScope~\cite{deng2018massive} and DCA~\cite{eraslan2019single} are based on autoencoders, obtaining effective feature representations through supervised learning. 
DCA introduces an autoencoder based on the zero-inflated negative binomial (ZINB) model to simulate the distribution of scRNA-seq data, generating a representation matrix to capture nonlinear dependencies between genes. 
SCVI~\cite{lopez2018deep} and SCVIS~\cite{ding2018interpretable}, which rely on autoencoders for dimensionality reduction, often face over-regularization due to their Gaussian distribution assumption. 
In contrast, DeepImpute~\cite{arisdakessian2019deepimpute}, based on standard deep neural networks, focuses on imputing missing values. 
DEC~\cite{xie2016unsupervised} and DESC~\cite{li2020deep} employ neural networks to learn feature representations and clustering assignments. DESC gradually eliminates batch effects through self-learning iterations, reducing technical differences between batches to approach real biological variations. 
However, these methods primarily focus on denoising and imputing dropout values in scRNA-seq data. 
Additionally, scDeepCluster~\cite{tian2019clustering} combines ZINB-based autoencoders with deep embedding clustering to optimize latent feature learning and clustering for better results. 
scDCC~\cite{tian2021model} encodes prior knowledge as constraint information and integrates it into the representation learning process through a novel loss function.  
AttentionAE-sc~\cite{li2023attention} combines denoising representation learning with cluster-friendly representation learning using attention mechanisms. 
However, the aforementioned approaches predominantly concentrate on extracting features from isolated cells, neglecting the vital intercellular structural information, crucial for accurately representing the variations between different cell types.

\subsection{Deep Structural Clustering Algorithm}
Recently, deep structural clustering methods have gained considerable attention and achieved remarkable effectiveness. 
scGAE~\cite{luo2021topology} and graph-sc~\cite{ciortan2022gnn} utilize graph autoencoders to embed scRNA-seq data while preserving their topological structures. 
scVGAE~\cite{inoue2024scvgae} integrates GNNs into a variational autoencoder framework and utilizes a ZINB loss function. 
scDeepSort~\cite{shao2021scdeepsort} pioneers the use of pre-trained GNN models for annotating cell types in scRNA-seq data and introduces a strategy using gene expression values to construct cell-gene graphs. 
scGAC~\cite{cheng2022scgac} constructs cell graphs and adopts a self-optimization approach for simultaneous learning of representations and clustering optimization. 
Notably, scGNN~\cite{wang2021scgnn} utilizes GNNs and multi-modal autoencoders to aggregate cell-cell relationships and model gene expression patterns using a left truncated mixture Gaussian model, scDSC~\cite{gan2022deep} combines a ZINB model-based autoencoder with GNN modules, employing a mutually supervised strategy for enhanced data representation. 
GraphSCC~\cite{zeng2020graph} employs GNNs to capture structural cell-cell relationships, optimizing the network’s output with a self-supervised module.
However, these GNN-based methods often face issues like representation collapse, and their high complexity and limited scalability often hinder their application to large-scale scRNA-seq datasets.

\newpage


\bibliographystyle{named}
\bibliography{reference}

\end{document}